\documentclass[twocolumn]{aastex6}
\usepackage{multirow}
\usepackage{amsbsy}
\usepackage{wasysym}
\usepackage{newfloat}

\DeclareFloatingEnvironment[name={Supplementary Figure}]{suppfigure}
\begin{document}

\title{The Gould's Belt Distances Survey (GOBELINS). V. Distances and Kinematics of the Perseus molecular cloud
%SD: wigh VLBA \& Gaia
}

\author{Gisela N.\ Ortiz-Le\'on\altaffilmark{1,14}, 
Laurent Loinard\altaffilmark{2,3}, 
Sergio A.\ Dzib\altaffilmark{1},
Phillip A.\ B.\ Galli\altaffilmark{4},
Marina Kounkel\altaffilmark{5},
Amy J.\ Mioduszewski\altaffilmark{6},
Luis F.\ Rodr\'{\i}guez\altaffilmark{2},
Rosa M.\ Torres\altaffilmark{7},
Lee Hartmann\altaffilmark{8},
Andrew F.\ Boden\altaffilmark{9},
Neal J.\ Evans II\altaffilmark{10},
Cesar Brice\~no\altaffilmark{11}, and
John J.\ Tobin\altaffilmark{12,13}
}

\email{gortiz@mpifr-bonn.mpg.de}

\altaffiltext{1}{Max Planck Institut f\"ur Radioastronomie, Auf dem H\"ugel 69, 
D-53121 Bonn, Germany}
\altaffiltext{2}{Instituto de Radioastronom\'ia y Astrof\'isica, 
Universidad Nacional Aut\'onoma de Mexico,
Morelia 58089, Mexico}
\altaffiltext{3}{Instituto de Astronom\'ia, Universidad Nacional Aut\'onoma de M\'exico, Apartado Postal 70-264, 04510 Ciudad de M\'exico, M\'exico}
\altaffiltext{4}{Instituto de Astronomia, Geof\'isica e Ci\^encias Atmosf\'ericas, Universidade de S\~ao Paulo,
Rua do Mat\~ao 1226, Cidade Universit\'aria, S\~ao Paulo, Brazil}
\altaffiltext{5}{Department of Physics and Astronomy, Western Washington University, 516 High St, Bellingham, WA 98225, USA}
\altaffiltext{6}{National Radio Astronomy Observatory, Domenici Science Operations 
Center, 1003 Lopezville Road, Socorro, NM 87801, USA}
\altaffiltext{7}{Centro Universitario de Tonal\'a, Universidad de Guadalajara,
Avenida Nuevo Perif\'erico No. 555, Ejido San Jos\'e
Tatepozco, C.P. 48525, Tonal\'a, Jalisco, M\'exico.}
\altaffiltext{8}{Department of Astronomy, University of Michigan, 500 Church Street, 
Ann Arbor, MI 48105,  USA}
\altaffiltext{9}{Division of Physics, Math and Astronomy, California Institute of Technology, 
1200 East California Boulevard, Pasadena, CA 91125, USA}
\altaffiltext{10}{Department of Astronomy, The University of Texas at Austin, 
2515 Speedway, Stop C1400, Austin, TX 78712-1205, USA}
\altaffiltext{11}{Cerro Tololo Interamerican Observatory, Casilla 603, La Serena, Chile}
\altaffiltext{12}{Homer L. Dodge Department of Physics and Astronomy, University of Oklahoma, 440 W. Brooks Street, Norman, OK 73019, USA}
\altaffiltext{13}{Leiden Observatory, PO Box 9513, NL-2300 RA, Leiden, The Netherlands}
\altaffiltext{14}{Humboldt Fellow}

\begin{abstract}
We derive the distance and structure of the Perseus molecular cloud by combining trigonometric parallaxes from Very Long Baseline Array (VLBA) observations, taken as part of the GOBELINS survey, and Gaia Data Release 2.  Based on our VLBA astrometry, we obtain a  distance of $321\pm10$~pc for IC~348. This is fully consistent with the mean distance of $320\pm26$ measured by Gaia. The VLBA observations toward NGC~1333 are insufficient to claim a successful distance measurement to this cluster. Gaia parallaxes, on the other hand, yield a mean distance of $293\pm22$~pc. Hence, the distance along the line of sight between the eastern and western edges of the cloud is $\sim$30~pc, which is significantly smaller than previously inferred.   We use Gaia proper motions and published radial velocities to derive the spatial velocities of a selected sample of stars. The average velocity vectors with respect to the LSR are $(\overline{u},\overline{v},\overline{w})$ = $(-6.1\pm1.6, 6.8\pm1.1, -0.9\pm1.2)$ and $(-6.4\pm1.0, 2.1\pm1.4, -2.4\pm1.0)$ km~s$^{-1}$ for IC~348 and NGC~1333, respectively. Finally, our analysis of the kinematics of the stars has shown that there is no clear evidence of expansion, contraction, or rotational motions within the clusters. 
%were taken in the Perseus molecular cloud .   

 \end{abstract}


\section{Introduction}\label{sec:intro}

The Perseus molecular cloud represents an ideal
%a well suited 
target for studying the fundamental properties of young stars and their environment, since the complex is sufficiently nearby   %relatively close 
that spatial scales down to $\sim$50 au are possible to reach with major observing facilities like ALMA and the VLA. Consisting of an elongated chain of dark clouds,  Perseus  spans over an area of $7^{\rm o}\times3^{\rm o}$ in the plane of the sky. The most prominent sub-structures are Barnard 5 (B5) and IC~348,  at the eastern edge, and Barnard 1 (B1), NGC~1333, L1448, L1451 and L1455 at the western edge  of the complex \citep[see e.g.][for a comprehensive review]{Bally2008}.
%(Herbst 2008).  
Most of the young stars reside in IC~348 and NGC~1333, which contains about 480 and 200 objects, respectively, with ages of $1-3$~Myr \citep{Luhman2016}, mainly identified from optical and near-IR surveys. The protostellar content within Perseus, on the other hand, has been probed with observations at mid-IR (Spitzer; \citealt{Enoch2009}), far-IR (Herschel; \citealt{Sadavoy2014}), sub-mm (JCMT; \citealt{Sadavoy2010}) and radio (VLA; \citealt{Tobin2016,Tychoniec2018}) wavelengths.  A total of  94 Class~0/I protostars and flat spectrum/Class~II objects are known to populate the entire cloud \citep{Tobin2016}.  

Multiple measurements of the distance to the individual clouds in Perseus have been performed in the past. 
%A short summary of these measurements is given by \cite{Zucker2018}. 
These measurements suggest that there is a distance gradient across the cloud, with  values in the range from 212 to 260~pc for the western component of the cloud \citep{Cernis1990,Lombardi2010,Hirota2008,Hirota2011,Schlafly2014} and 260--315~pc for the eastern component \citep{Cernis1993,Lombardi2010,Schlafly2014}. Direct measurement of distances via the trigonometric parallax have been  obtained for only a few sources in these regions. Based on Very Long Baseline Interferometry (VLBI) observations of H$_2$O masers associated with two YSOs in NGC~1333 and L1448, \cite{Hirota2008,Hirota2011} found a distance consistent with 235~pc for both clouds. 
%Whether Perseus consists of clouds at several distances or  
However, whether or not the gradient in the distance across the whole complex is significant remains inconclusive, since the distance uncertainties on individual lines of sight are large (typically $\sim10-20\%$ for photometric distances), and the number of sources with available direct distance measurements is small. 

In the past few years, we have used the Very Long Baseline Array (VLBA) to measure the trigonometric parallax of several tens of 
%almost one hundred
young stars in nearby star-forming regions \citep[][]{Ortiz2017Oph,Kounkel2017,Ortiz2017Ser,Galli2018} as part of the Gould's Belt Distances Survey (GOBELINS) project. Very Long Baseline Interferometry (VLBI) has the advantage of being able to detect highly embedded sources, where the  extinction by dust obscures the optical light from the stellar objects. Given the high angular resolution provided by the VLBA and the fact that the interstellar material in these regions is transparent to radio waves, parallaxes with an accuracy of  $1\%$ or better are possible for these kind of sources.  In addition, parallaxes toward more than four hundred stars in Perseus with a limiting magnitude G=21~mag and parallax uncertainties $<0.7~$mas  have become available during the second Gaia data release (DR2).  With this highly  accurate astrometric data, we can now investigate the depth of the molecular cloud and the three-dimensional motions of the young stars as well as the global properties of the kinematics of IC~348 and NGC~1333. 

We first describe the VLBA observations in Section \ref{sec:obs} and the fits to our data in Section \ref{sec:astro-vlba}. Section \ref{sec:gaia} presents the extraction of the astrometric solutions from the Gaia DR2 catalog. We then use both VLBA and Gaia data to investigate the structure of the Perseus cloud, which is discussed in Section \ref{sec:structure}. Sections \ref{sec:pm} and \ref{sec:vel} present the kinematics of a selected sample of cluster members in IC~348 and NGC~1333. Finally, our conclusions are given in Section \ref{sec:conclusions}.  

\section{VLBA Observations and data reduction}\label{sec:obs}

The target selection for the VLBA survey and observing strategy follows the same procedure described in detail in \cite{Ortiz2017Oph}. In summary, we constructed our target sample based on the properties of the radio emission detected with the Very Large Array toward Young stellar objects (YSOs) and YSO candidates   in NGC~1333 and IC~348 \citep{Pech2016}. We observed all radio sources associated with YSOs whose radio emission could be detected with the VLBA, i.e.\ non-thermal sources. In addition, we observed all unidentified sources in the region whose radio properties are consistent with YSOs and  have fluxes above the threshold of the GOBELINS observations. In total, 59 sources were observed between April 2011 and March 2018 at $\nu=5.0$ or 8.4~GHz (C- and X-band, respectively). The data were recorded in dual polarization mode with 256 MHz of bandwidth in each polarization, covered by eight separate 32 MHz intermediate frequency (IF) channels.  Each observing session consisted of  cycles alternating between the target and J0336+3218. Three additional calibrators were observed every $\sim50$~minutes to improve the phase calibration. In addition, geodetic-like blocks, consisting of observations of many calibrators over a wide range of elevations, were taken before and after each session. We use AIPS \citep{Greisen2003} for data inspection,  calibration and imaging, following standard procedures for phase-referencing observations as described in \cite{Ortiz2017Oph}. 

Out of the 25 sources detected, only 7 are related to YSOs, while the rest turned out to be background objects with negligible motion on the plane of the sky.  In this paper, we present a subset of detected YSOs for which we can measure both parallax  and proper motions (4 sources in total). The other 3 sources have detections in only 1-2 epochs, which is insufficient to perform the astrometric fits. The dates of these observations and VLBA pointing positions are given in Table \ref{tab:obs}. 

\section{VLBA Astrometry}\label{sec:astro-vlba}

Source positions at individual epochs were extracted by performing two-dimensional Gaussian fits with the AIPS task JMFIT (Table \ref{tab:positions}).  Parallax, $\varpi$, position at median epoch, $(\alpha_0,\delta_0)$,  and proper motions, $\mu_\alpha$ and $\mu_\delta$, were fitted to the measured positions by minimizing $\chi^2$ in each direction. %The fitted parameters are given in Table \ref{tab:gaia-prlx}. 
Systematic errors were added to the statistical errors provided by JMFIT. These errors were  obtained by scaling positional uncertainties until the reduced $\chi^2$ of the fit becomes equal to 1. The resulting best-fit parameters are shown in columns (2)-(4) in Table \ref{tab:gaia-prlx}. We briefly discuss each source in the following paragraphs. 

\subsection{IRAS 03260+3111 = 2MASS J03291037+3121591 }

This is a known wide binary system with a separation of 3.62$''$ \citep{Haisch2004} located in NGC~1333. It has been classified as a Class~II object \citep{Gutermuth2008} of spectral type F5 \citep{Luhman2016}. We  simultaneously detected  two sources  with the VLBA in two epochs, with an angular separation of $\sim0.7''$. The  closer companion was already seen by \cite{Connelley2008}, who found a binary separation of 0.55$''$ in the near-IR.  The source is thus a  hierarchical triple system. The brightest IR component corresponds to the western radio source seen in our maps. 

%The data are  not enough to fit the orbital parameters. 
Given the scarcity of the data from the only 5 epochs available, the fit including orbital motions does not converge to reliable parameters, and the corresponding uncertainties are large. We thus fit only parallax and proper motions separately to each source, and adopt the resulting parameters from the fit to the eastern component, which has 4 detected epochs. 

\subsection{V913 Per} 

This source is located in IC~348. It is a Class~III star of spectral type M2.5  \citep{Luhman2016}. It has been detected in 7 epochs, which we used for the derivation of the astrometric parameters.  The derived parallax has an uncertainty  of 3.3\%. 

\begin{figure*}[!ht]
\begin{center}
\includegraphics[width=0.5\textwidth,angle=0]{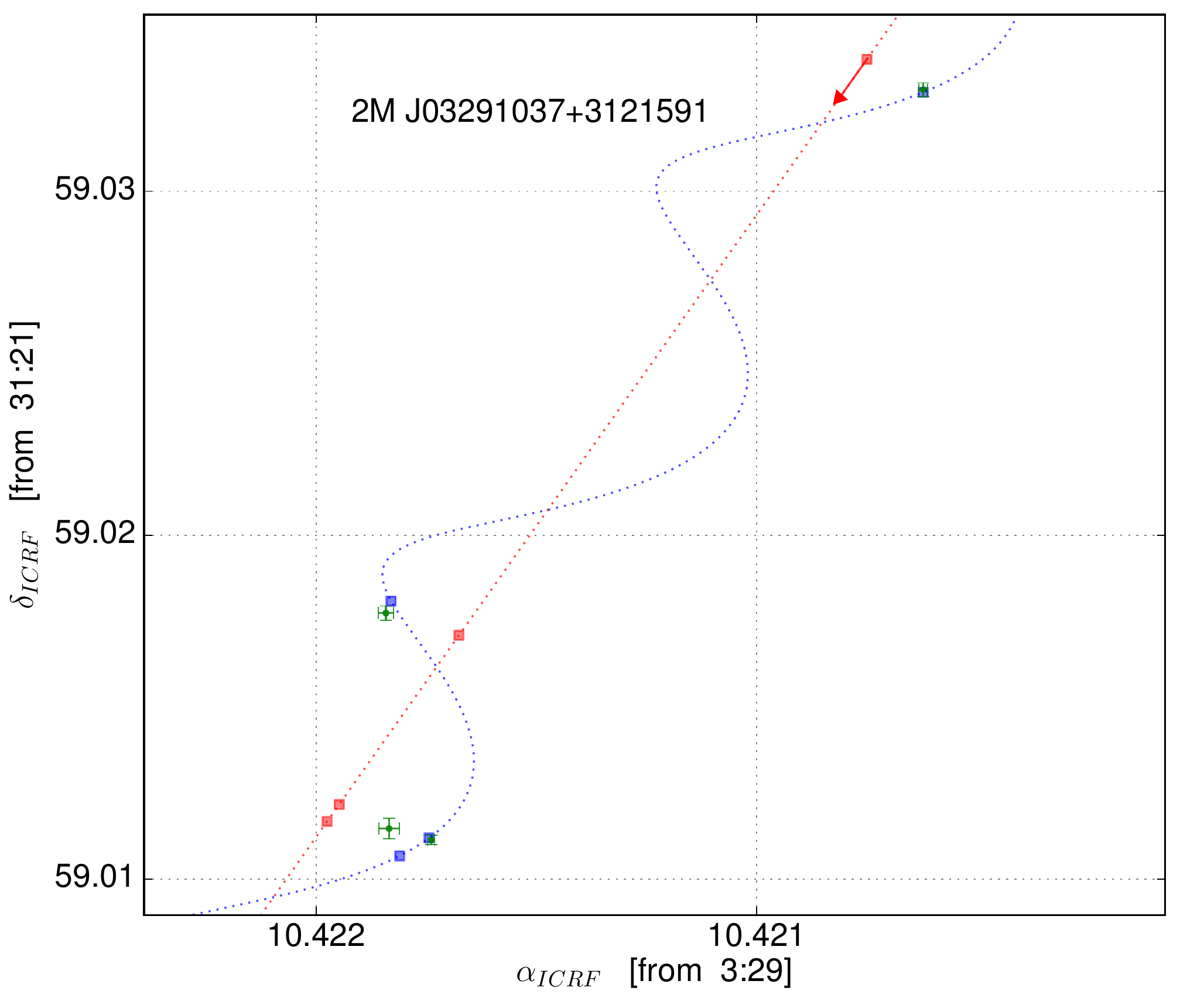} \\ 
 \includegraphics[width=0.4\textwidth,angle=0]{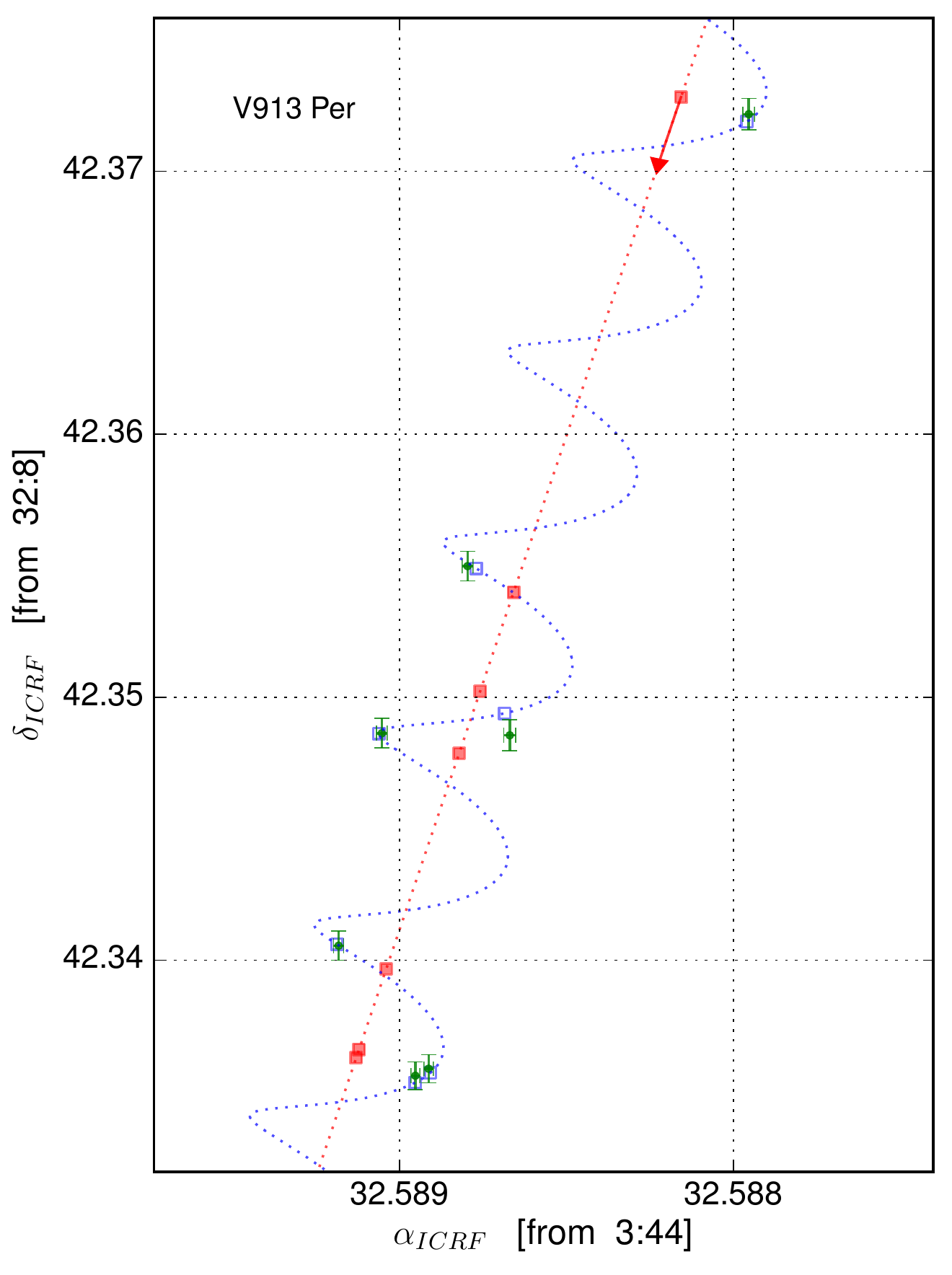}
  \includegraphics[width=0.39\textwidth,angle=0]{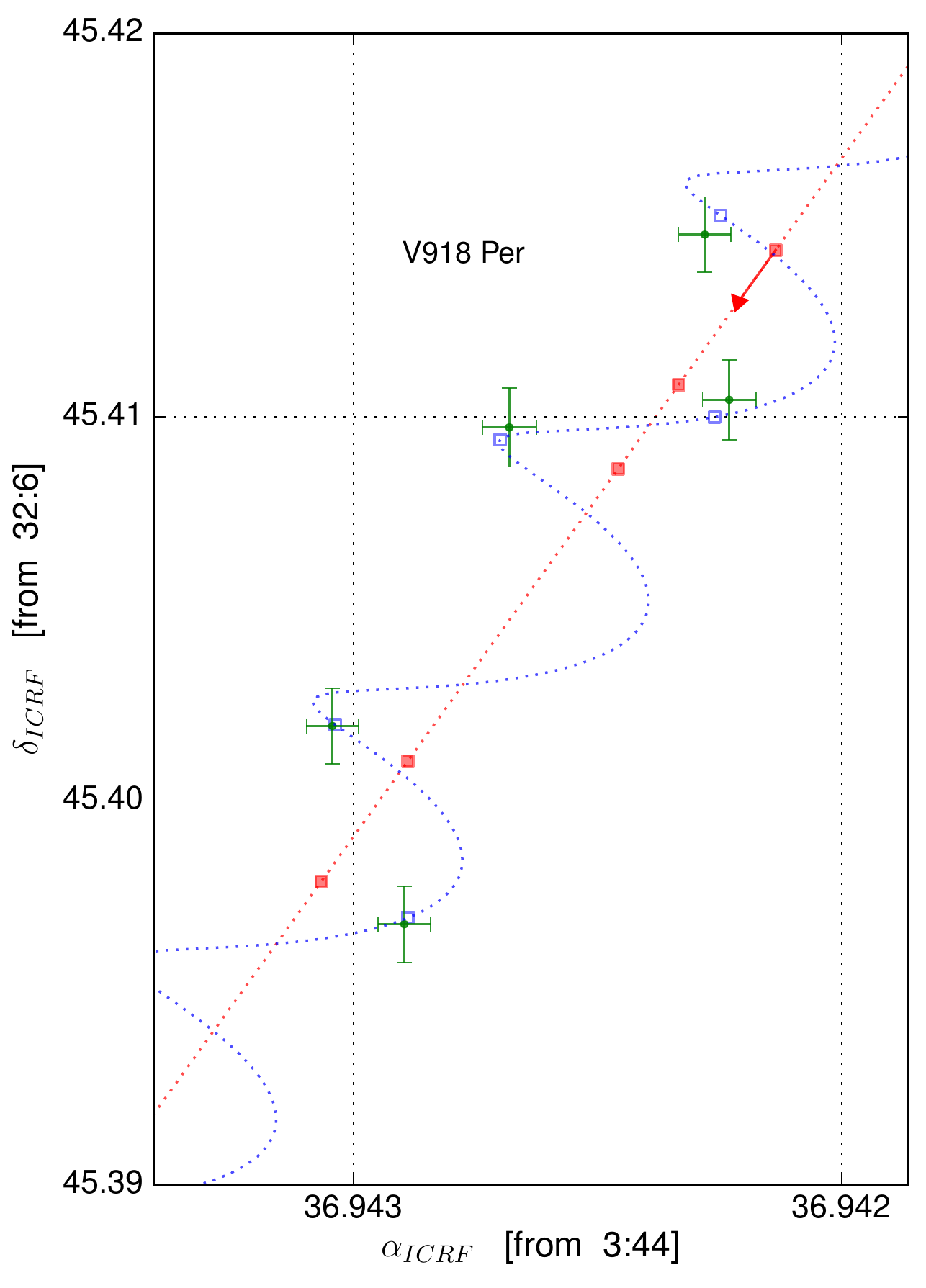}
\end{center}
 \caption{Astrometric fits to VLBA data. Observed positions and expected positions from the fits are shown as green dots and blue open squares, respectively. The blue dotted line is the fitted model, and the red line is the model with the parallax signature removed. The red squares mark the position of the sources from the model without parallax. The arrows show the the direction of position change over time. }
\label{fig:v913}
\end{figure*}

\subsection{V918 Per} 

Also located in IC~348, it is a Class~II/III object \citep{Alexander2012,Kaisa2015} with a spectral type of G3 \citep{Luhman2016}. Two sources have been detected in our maps in alternative epochs. One source was only seen  in the first  epoch, while the other source has been detected in 5 epochs.  We fit only the astrometric parameters to these five epochs.  
% Offset between components is 0.551 arcsec
% Radial velocity is 20.021965000 km/s $\pm$ 0.5785095692 
%     RASHIFT       0.2438, 0; DECSHIFT     -0.4578, 0 

\subsection{LRL 11} 

This source is a Class~III star with a spectral type of G4 \citep{Luhman2016} located in IC~348 as well.  The model including only parallax and proper motions produces a poor fit to the data. We investigate if the source motion can be reproduced by adding an orbital component due to the possibility that the source is a binary system. The fit that includes orbital motions does indeed reproduce the measured source positions. We found that the two methods we have used in the past to fit  binaries \citep[c.f.][]{Kounkel2017,Galli2018} yield different solutions for the orbital elements.   This means that our data is not good enough to constrain the orbit, so it should be taken somewhat cautiously.  On the other hand, the parallax and proper motions from the two methods agree within $2\sigma$.  We give in Table \ref{tab:lrl11} the best-fit solution obtained from the MCMC method (Galli et al.\ 2018). 

\begin{figure*}[!ht]
\begin{center}
\includegraphics[width=0.5\textwidth,angle=0]{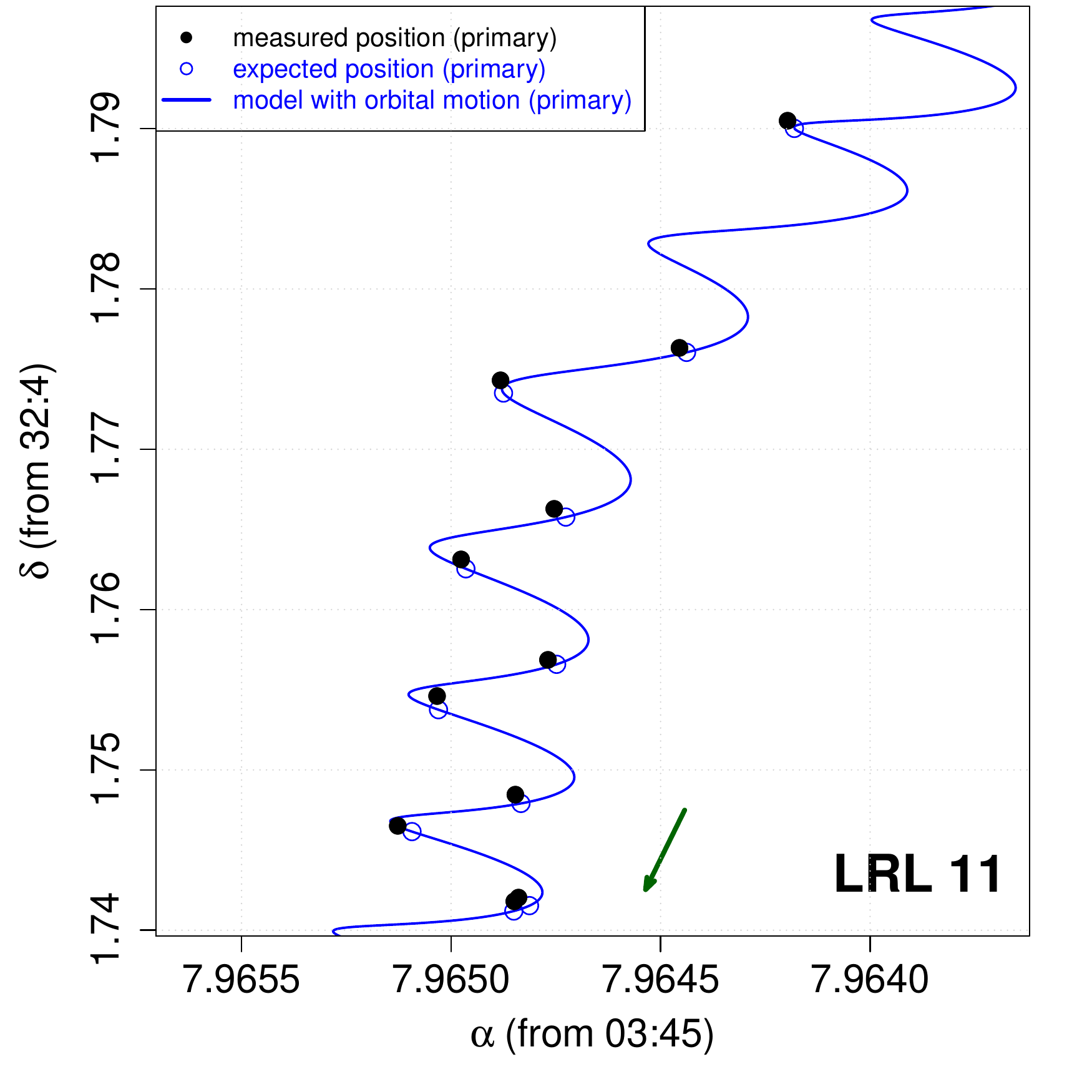}  
\end{center}
 \caption{Astrometric fit to VLBA data taken toward LRL~11 including orbital motion.  Observed positions and expected positions from the fits are shown as black filled  and blue open dots, respectively.  The blue line is the fitted model. }
\label{fig:lrl11}
\end{figure*}

\bigskip

\section{Gaia data}\label{sec:gaia}

With the recent release of Gaia DR2 \citep{gaia_2016,gaia_2018,Lindegren_2018}, astrometric data for objects with G$<$21~mag have become available. We will use this  wealth of new data to assess further the distance to Perseus. 

As discussed in Section \ref{sec:intro}, the most complete catalog to date of young members in IC~348 and  NGC~1333  has been  compiled by  \cite{Luhman2016}. This catalog contains 478 and 203  stars in IC~348 and  NGC~1333, respectively, for which memberships were confirmed from optical and near-IR spectroscopy. We performed a cross-match of the young stars positions against the Gaia DR2 catalog using a search radius of 1$''$. 
The source coordinates in the catalog of \cite{Luhman2016} were either taken from the 2MASS Point Source Catalog (which have a positional accuracy of $0.1-0.3"$) or measured from previous photometric infrared surveys, where allowed positional shifts are up to $\sim1"$ \citep{Alves2013}. The radial velocity catalogs we will use in Section \ref{sec:vel} have been constructed from a variety of published  X-ray, optical and mid-infrared surveys, where positional errors range from 0.3 to $1"$. Thus, the choice of a match radius of $1"$  allows us to take into account the different uncertainties from the various surveys. In total, 351 and 90 stars, in IC~348 and  NGC~1333, respectively, have five astrometric parameter solutions.

Three of our VLBA-detected sources appear in the Gaia DR2 catalog. Their astrometric solutions are given in columns (5)-(7) in Table \ref{tab:gaia-prlx}. The difference between VLBA and Gaia parallaxes is 0.589,  1.277 and 0.015~mas, for V913~Per, V918~Per and LRL~11. The proper motions are remarkably different mainly  in the R.A.\ direction. This discrepancy is expected for the binary systems, V918~Per and LRL~11, since all Gaia sources have been treated as single stars in DR2. We argue that the discrepancy in the astrometric solutions for V913~Per can be attributed to systematic errors present in Gaia DR2. The magnitude of these systematic errors is $\sim$0.1~mas for parallaxes and $\sim$0.1~mas~yr$^{-1}$ for proper motions \citep{Luri_2018}. In addition, a parallax zeropoint offset of $-0.03$~mas, corresponding to the mean parallax of sources identified as  quasars, should be also taken into account \citep{Lindegren_2018}. If the  systematic errors are added quadratically to the quoted uncertainties in Gaia DR2 catalog, then the parallax of V913~Per agrees within $2\sigma$, and the proper motion in declination within $1\sigma$. However, the proper motion in right ascension still disagrees by $\sim5\sigma$. For this particular source, the quantities {\it astrometric\_excess\_noise} and {\it astrometric\_excess\_noise\_sig}, given in the Gaia archive, have values of 1.1~mas and 187.6, respectively. These parameters represent the excess noise of the source and its significance, which measure the difference between the observations and the best-fitting astrometric model. Values of {\it astrometric\_excess\_noise} $>0$~mas (with {\it astrometric\_excess\_noise\_sig} $>$ 2) indicate that the residuals of the fit to the Gaia data are larger than expected due to modelling and calibration errors. Other VLBA sources we have monitored in Orion and Taurus  \citep[][Galli et al.\ 2018]{Kounkel2017} show an agreement in proper motion better than $2\sigma$. It is doubtful that our VLBA data are affected by systematic effects. 
%We thus discard the possibility that our VLBA data are affected by systematic effects. 

\begin{figure*}[!ht]
\begin{center}
\includegraphics[width=1.0\textwidth,angle=0]{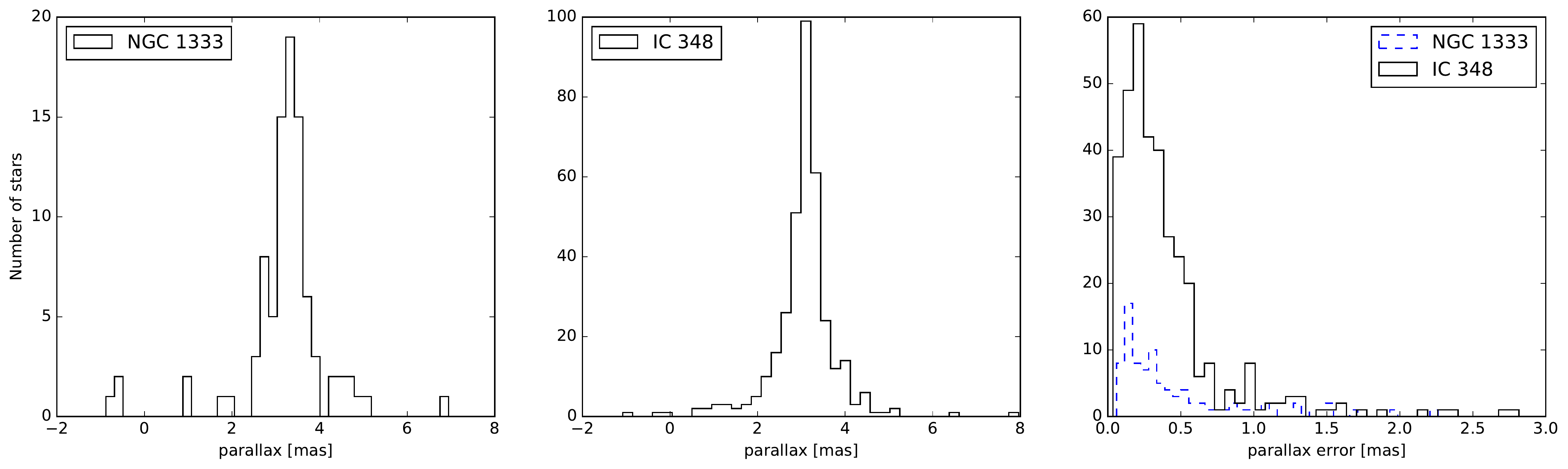}  
\end{center}
 \caption{Distributions of parallaxes and their uncertainties measured by Gaia towards Perseus. }
\label{fig:hist}
\end{figure*}

\section{Structure of Perseus}\label{sec:structure}

We show in Figure \ref{fig:hist} the distributions of  Gaia parallaxes and parallax uncertainties for NGC~1333 and IC~348. 
%, for which median   uncertainties are 0.30~mas. % in both NGC~1333 and IC~348. 
The parallax uncertainties have median values of 0.30~mas in both cases. 
The weighted mean of these parallaxes is $\varpi = 3.38\pm0.02$~mas with a weighted standard deviation of $\sigma_\varpi=0.32$~mas for NGC~1333, and  $\varpi=3.09\pm0.01$~mas with $\sigma_\varpi=0.26$~mas, for IC~348. Inverting the weighted mean parallaxes (after correcting for the parallax zero-point shift of $-30$~$\mu$as) yields a distance of $d = 294$~pc with a standard deviation of $\sigma_d = 28$~pc, for NGC~1333, and $d = 321$~pc with $\sigma_d = 27$~pc  for IC~348. If we remove the stars with parallaxes outside the core of the parallax distribution (i.e., with $\varpi<1.5$ and $>6$~mas in NGC~1333 and $\varpi<0.4$ and $>6$~mas in IC~348), the mean parallaxes give $d = 293\pm22$~pc for NGC~1333 and $d = 320\pm26$~pc for IC~348, where the quoted errors correspond to the standard deviation. In each cluster, the data have median of parallax uncertainties larger than  the standard deviation of the whole distribution.
%standard deviation 
This means that the parallax dispersion is not dominated by the intrinsic dispersion, but by the 
uncertainties on individual parallaxes. Thus, the true depth of the clouds cannot be extracted from these measurements. 

The VLBA data alone suggest that NGC~1333 and IC~348 are located at similar distances. However, only the source IRAS~03260+3111 in NGC~1333 was used for the present analysis. This source is a multiple system where the angular separation between the VLBA components is $\sim0.7"$. We do not expect that at such separation the orbital motion has a measurable effect on the motion of individual components over a time scale of a few years. The fit does not completely agree with the observations,  though (top panel in Figure \ref{fig:v913}), so it is still possible that an additional and much closer unseen companion is present in the system.
%Still more VLBA data would help to confirm the measurement. 
%where the orbital motion of the closer binary has not yet been taken into account when fitting the parallax. 
Regarding IC~348, the weighted mean of the VLBA parallaxes measured for  V913~Per and  V918~Per yields  $321\pm10$~pc. This is in agreement with the weighted mean distance derived  from  Gaia parallaxes. We thus recommend to use $321\pm10$~pc as the distance to IC~348 and $293\pm22$~pc for NGC~1333. 
% . 
Finally, we note that the binary system LRL~11, located at a distance of $373\pm11$~pc, may not be part of Perseus but a background object projected in the direction of IC~348. 

\begin{figure*}[!ht]
\begin{center}
\includegraphics[width=0.9\textwidth,angle=0]{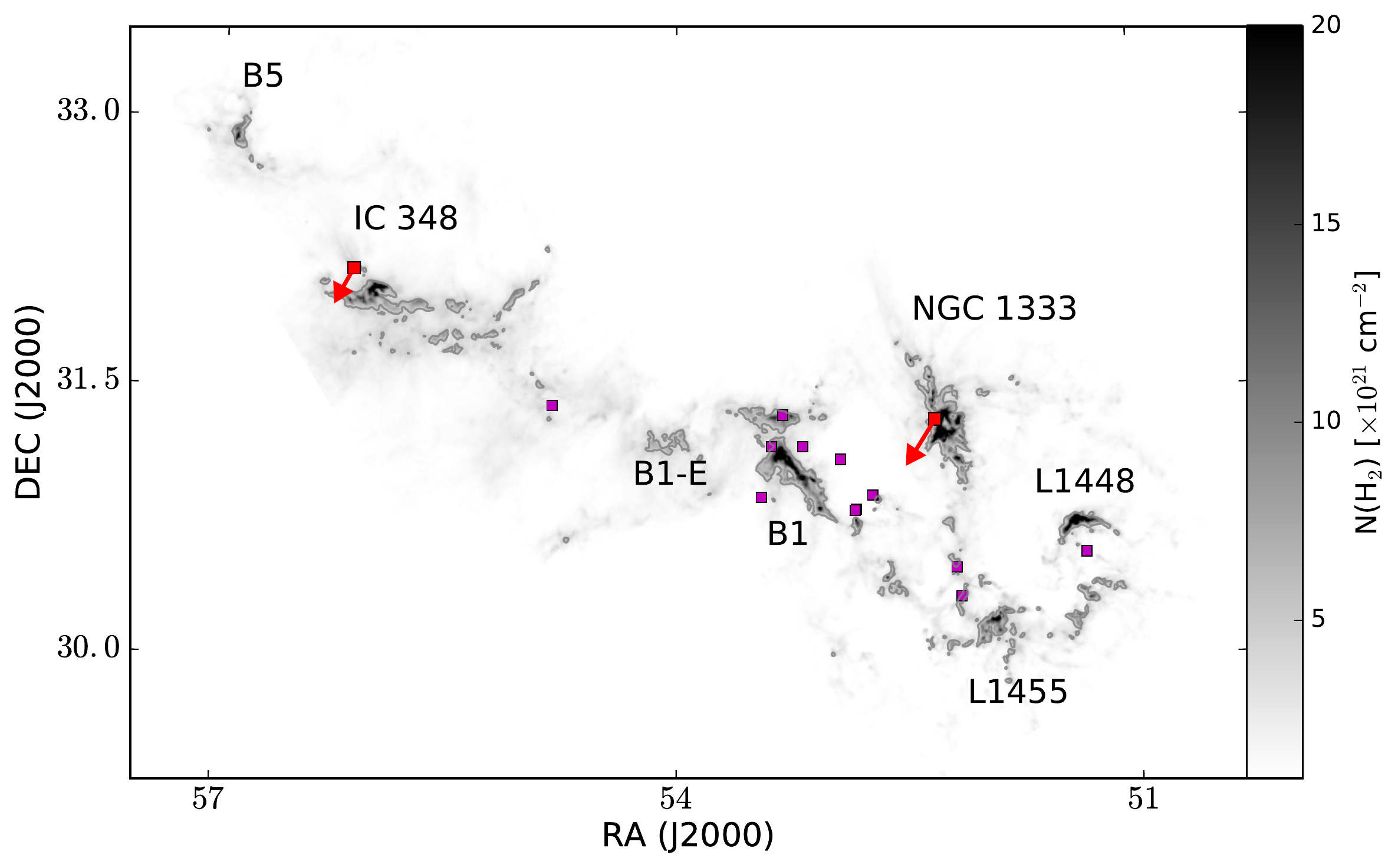}  
\end{center}
 \caption{Large scale column density map of the Perseus Molecular cloud from the {\it Herschel} Gould Belt survey \citep{Andre2010,Sadavoy2014}. The red arrows show the mean of the measured proper motions in IC~348 and NGC~1333. The origin of the arrows is at the mean position of the selected sample of stars described in Section \ref{sec:pm}.  The magenta squares are other YSOs candidates across Perseus with five astrometric solutions in the Gaia DR2 catalog. }
\label{fig:perseus}
\end{figure*}

Outside IC~348 and NGC~1333, we found Gaia parallaxes for 13 objects (see Figure \ref{fig:perseus} and Table \ref{tab:other-gaia}) which are known YSO candidates in Perseus \citep{Dunham_2015}. One object resides in the outskirts of L1448 and has a parallax of $0.68\pm0.43$~mas, which means it is not part of Perseus. Two objects are in the outskirts of L1455, with parallaxes of $1.90\pm1.98$~mas and $3.62\pm0.14$~mas, respectively. While the former value does not provide any useful information due to its large  uncertainty, the latter is consistent within 2$\sigma$ with the mean of parallaxes measured in NGC~1333. Other 9 objects are projected in the direction of the cloud B1. They have a weighted mean parallax of $3.35\pm0.06$~mas with standard deviation 0.32~mas (corresponding to $296\pm$28~pc), which is also consistent with the weighted mean parallax  of NGC~1333 . The last object is found southwest of IC~348 and has a parallax of $2.37\pm0.70$~mas. The large uncertainty of this measurement makes it difficult to claim the actual distance to this star. 

Putting it all together, both Gaia and VLBA measurements suggest that the eastern edge of Perseus could be about 28~pc  farther than the western edge, which is a significantly smaller distance variation than previously thought \citep[e.g.][]{Hirota2011}. Past measurements of parallaxes using also VLBI resulted in a distance of $235\pm18$~pc for NGC~1333 \citep{Hirota2011}. The difference between this measurement and the derived in this work is $2.6\sigma$. We should note that the measurements by \cite{Hirota2011} were obtained from a fit to water masers, whose flux and positions showed time variability during the observing period of 6 months.  The peak velocity of the water emission also suffered a drift of $\sim0.9$~km~s$^{-1}$. These authors took the average position of the different maser ``spots'' (emission seen in a single velocity channel) over contiguous spectral channels, which resulted in two spatially separated ``features'' detected each during 3 and 4.5 months, respectively.  We discussed in \cite{Dzib2018} that this approach can introduce position fluctuations larger than the synthesized beam size, and introduce additional uncertainty in the astrometric parameters of the masers.  We demonstrated in \cite{Dzib2018} that, in order to reduce the chances of misidentifiying maser spots from one epoch to another, one should fit the maser positions measured at the same velocity channel in all epochs. In L1448, \cite{Hirota2011} also detected several spots at different velocity channels. In this case, the authors did fit only maser spots detected at a velocity of $\sim20.6$~km~s$^{-1}$. However, their data span a baseline of only 5 months, which is not enough to properly cover the parallax sinusoid. It is possible that the variability  of the maser emission led to a misidentification of the maser spots and affected the astrometry performed toward these masers. Unfortunately, the protostars to which these masers are associated  are too embedded that they remained undetected by Gaia, so a direct comparison against Gaia astrometry is not possible at the moment. It is also possible that NGC~1333 has multiple components  along the line of sight. However, as we already mention above, the Gaia parallaxes are not good enough to search for such components. 

\section{Kinematics of IC~348 and NGC~1333}
\subsection{Proper motions} \label{sec:pm}

\begin{figure*}[!ht]
\begin{center}
\includegraphics[width=0.7\textwidth,angle=0]{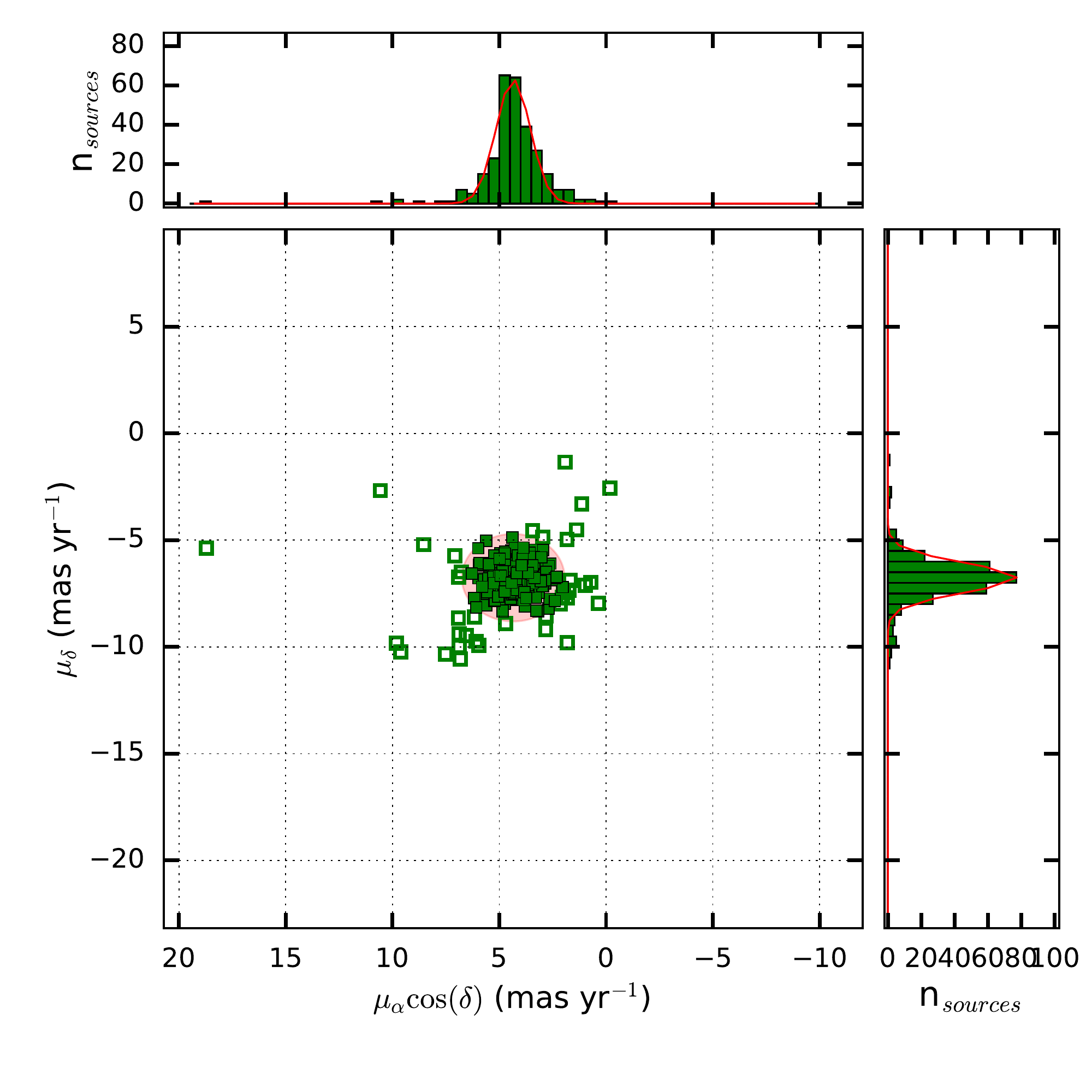}  
\end{center}
 \caption{Proper motions measured by Gaia in IC~348. The top and right panels show the distributions of $\mu_\alpha\cos\delta$ and $\mu_\delta$, respectively. The Gaussian fits to these distributions are plotted in red.  The filled and open squares are stars with proper motions within and outside $\pm3\sigma$ from the mean, respectively. The $\pm3\sigma$  range in proper motions is covered by the red shadow. 
 }
\label{fig:ic348-pm-hist}
\end{figure*}

\begin{figure*}[!ht]
\begin{center}
\includegraphics[width=0.7\textwidth,angle=0]{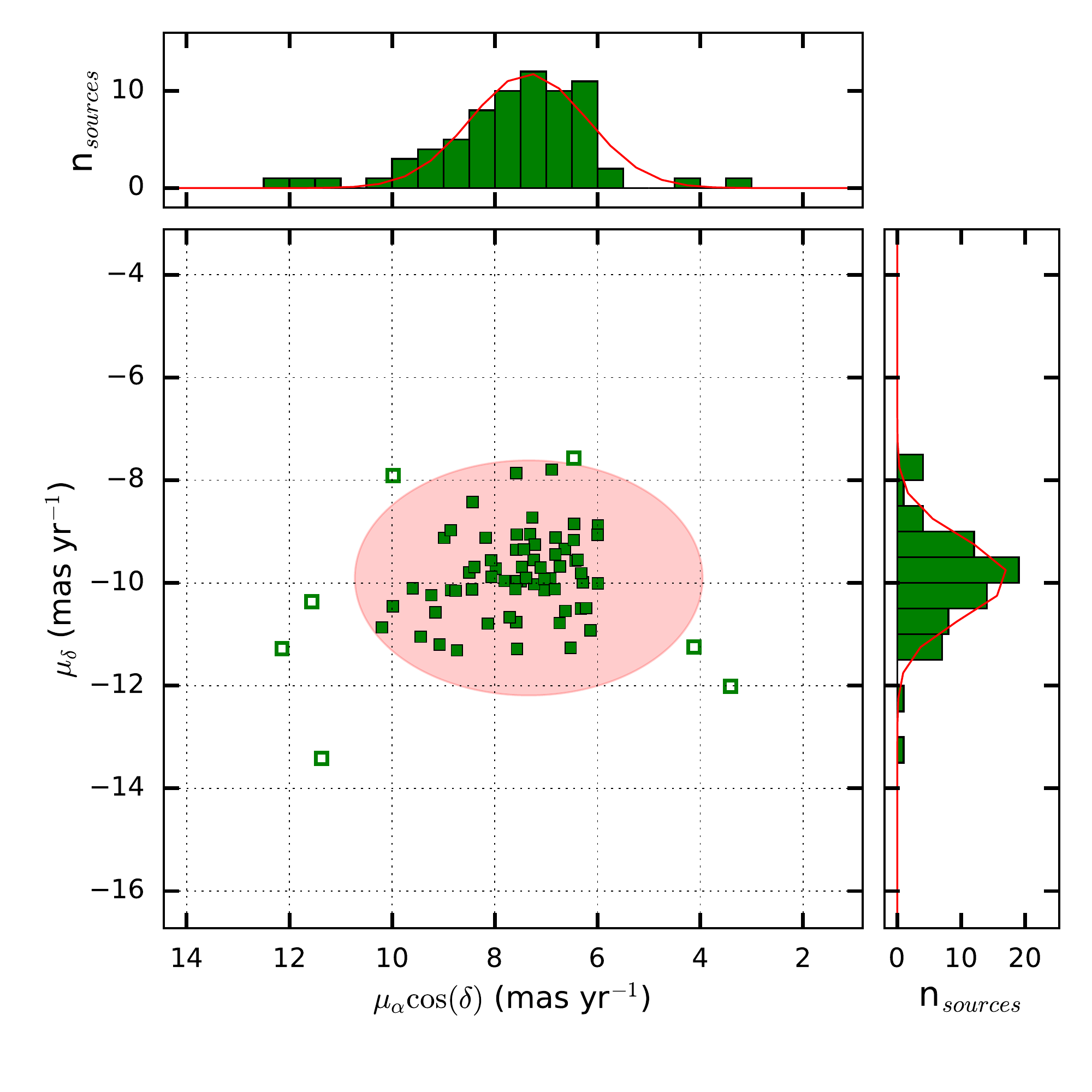}  
\end{center}
 \caption{Same as Figure \ref{fig:ic348-pm-hist} but for NGC~1333. }
\label{fig:ngc1333-pm-hist}
\end{figure*}

To analyse the proper motions within NGC~1333 and IC~348 and their intrinsic velocity dispersion, we need first to define a subset of cluster members which reflect the true dynamics of the clusters.  To construct such a sample, we exclude all sources with parallaxes that deviate by more than $3\sigma$ from the weighted mean parallax in each cluster. The distributions of  measured proper motions of the resulting sample after this initial cut are shown in Figures \ref{fig:ic348-pm-hist} and  \ref{fig:ngc1333-pm-hist}, for IC~348 and NGC~1333, respectively.  The proper motion distributions  were then fitted with Gaussian models, which are also plotted in red in these figures. We give in Table \ref{tab:properMotions} the mean and  velocity dispersion (corrected for the measurement errors) that result from the best-fit Gaussian distributions.  To  convert  proper motion dispersions into tangential velocity dispersions, we have used the mean distance of $321 \pm 10$~pc for IC~348, and $293 \pm 22$~pc for NGC~1333. 
Based solely on the analysis of the radial velocity distribution, \cite{Cottaar2015} measured a velocity dispersion of  $0.72 \pm 0.07$~km~s$^{-1}$ for IC~348. Similarly, \cite{Foster2015} measured $0.92\pm0.12$~km~s$^{-1}$ for NGC~1333. These values are comparable to the velocity dispersion measured here for the proper motions. 

\begin{figure*}[!ht]
\begin{center}
\includegraphics[width=0.52\textwidth,angle=0]{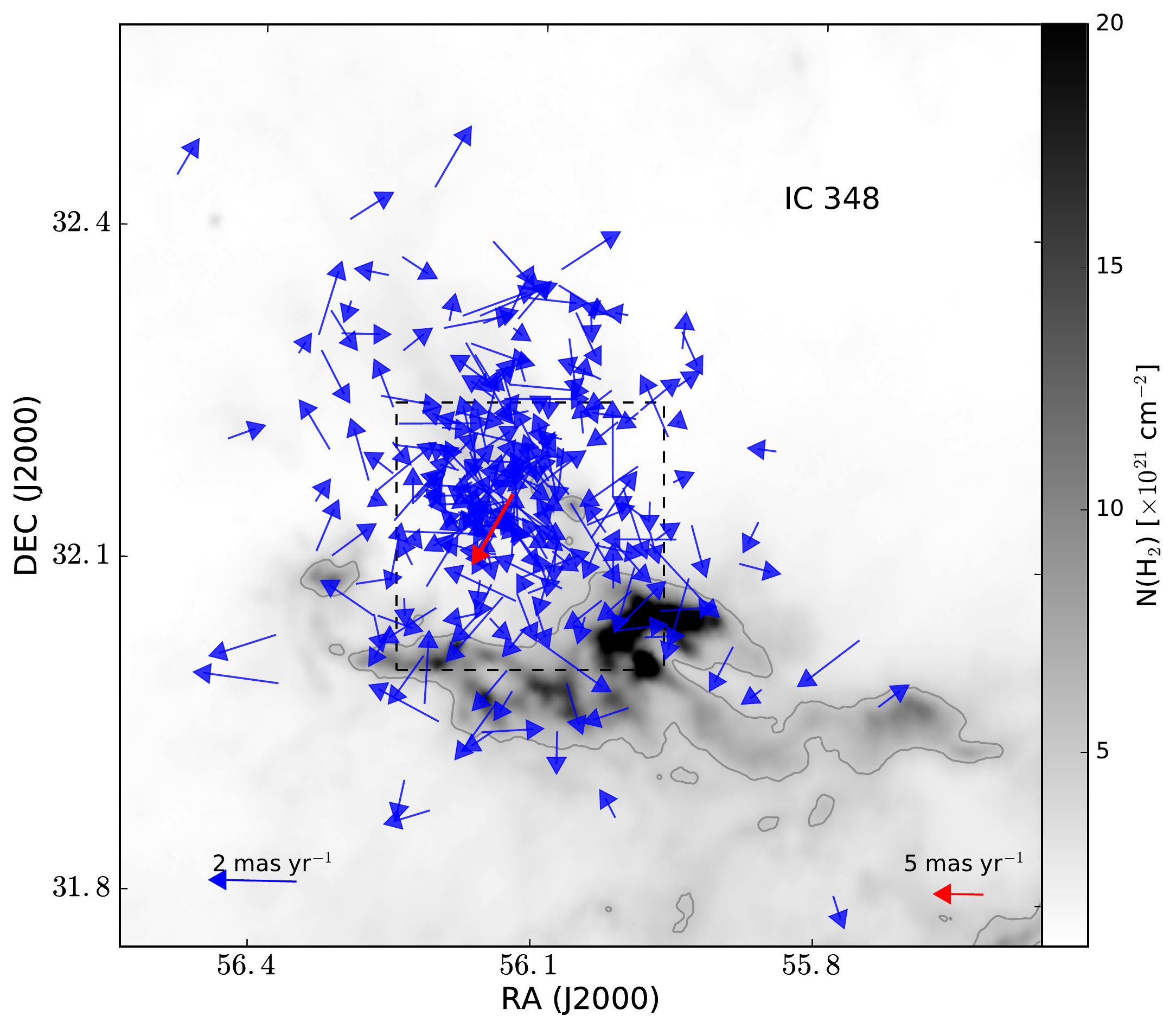}  
\includegraphics[width=0.47\textwidth,angle=0]{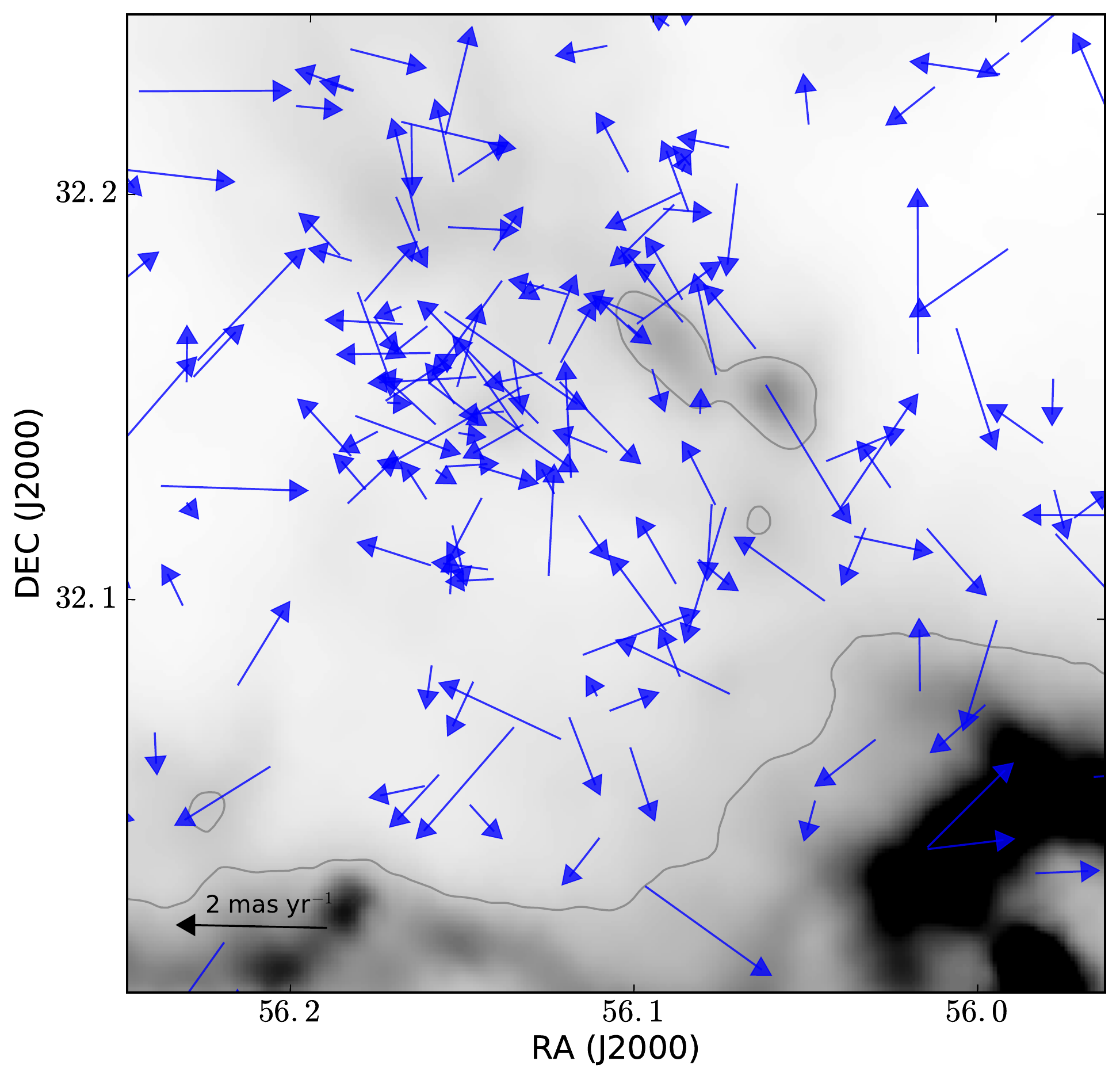}  
\end{center}
 \caption{Measured proper motions by Gaia overlaid on the column density map derived from the {\it Herschel} Gould Belt survey \citep{Andre2010} data  by \cite{Sadavoy2014}. The contour corresponds to $N({\rm H}_2)=5\times10^{21}$~cm$^{-2}$. The red arrow indicates the mean proper motion of the cluster obtained from a Gaussian fit to a subset of stars as described in Section \ref{sec:pm}. The blue arrows are individual measurements after subtracting the mean proper motion. The right panel shows a zoom-in of the central part of the left panel (dashed square).}
\label{fig:ic348-vel}
\end{figure*}

\begin{figure*}[!ht]
\begin{center}
\includegraphics[width=0.6\textwidth,angle=0]{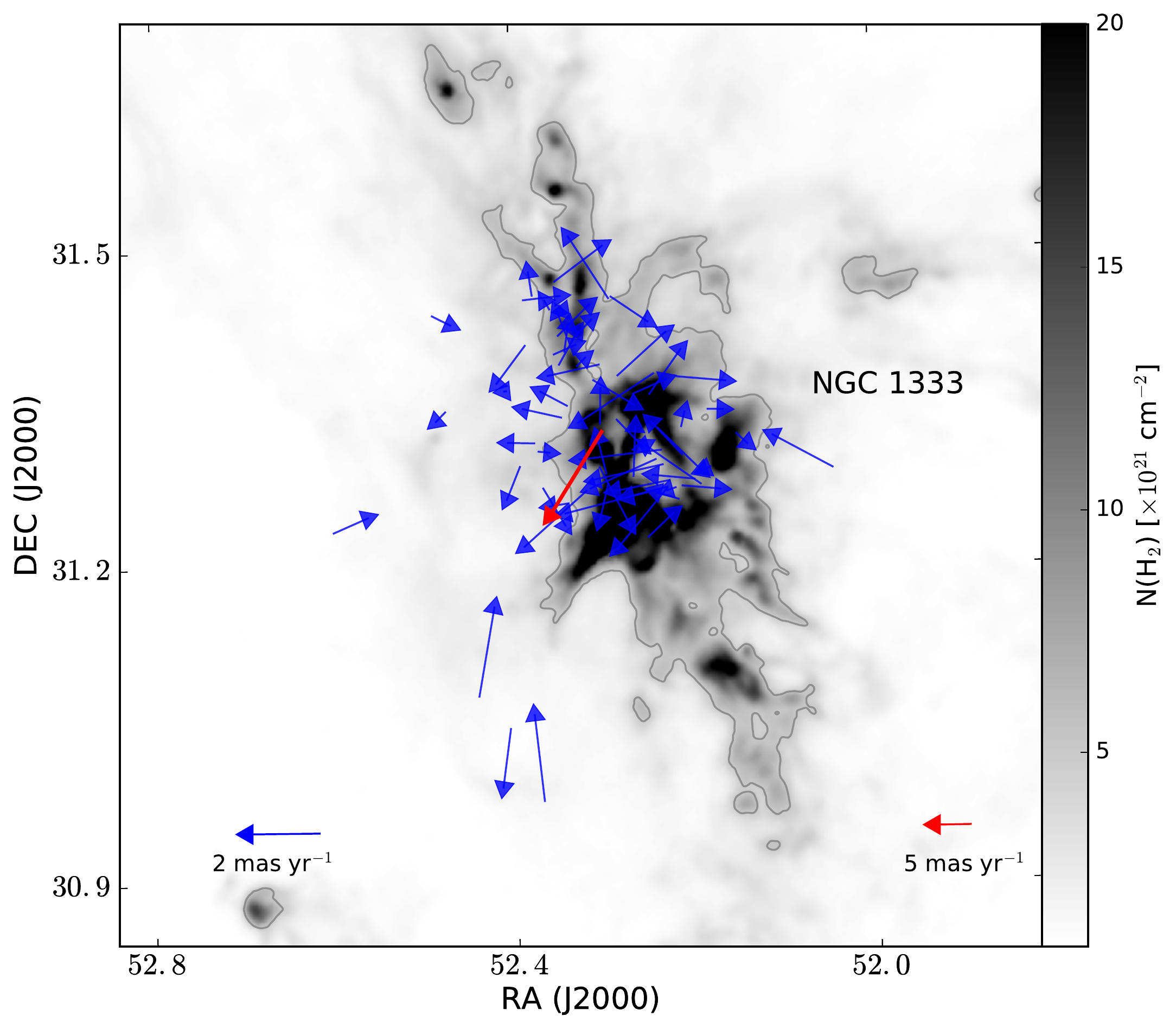}  
\end{center}
 \caption{Same as Figure \ref{fig:ic348-vel}, but for NGC~1333. }
\label{fig:ngc1333-vel}
\end{figure*}

We then cut further  stars with proper motions outside  $\pm3\sigma$ from the mean, where $\sigma$ is the measured standard deviation.  This selection has been made to mitigate the effects of unresolved astrometric binaries within our samples. Orbital motions are expected to contribute to the dispersion of the proper motions distributions in a non-preferential orientation. In Figure \ref{fig:ic348-pm-hist}, the stars that are cut by this criteria are plotted as green empty squares, while the green filled squares represent the clusters members used in the forthcoming analysis.   The proper motions of these subsets, relative to the mean of each cluster, are displayed in Figures \ref{fig:ic348-vel} and \ref{fig:ngc1333-vel}, while the measured values are listed in Table \ref{tab:astro}. We see that proper motions within each cluster are highly consistent between themselves, with mean magnitudes indicated by the red arrows in each Figure and given in Table \ref{tab:properMotions}.  Because these proper motions are measured relative to Sun, they mostly trace the reflex motion of the Sun.  We must thus remove the Solar motion for the analysis of  the internal kinematics of the stars in Perseus. 

% to minimize the effect of binary orbital motions

%yields mean velocities of $\bar{\mu}_\alpha\cos\delta = 3.65 \pm 0.02$~mas~yr$^{-1}$, and $\bar{\mu}_\delta = -6.76 \pm 0.01$~mas~yr$^{-1}$, with measured $\sigma_{\mu_\alpha\cos\delta}=0.84$~mas~yr$^{-1}$ and  $\sigma_{\mu_\delta}=0.73$~mas~yr$^{-1}$. 

\subsection{Spatial velocities}\label{sec:vel}

We now compute the three dimensional Galactic spatial velocities of the reduced sample of stars described in the section above. This requires the conversion of  proper motions and radial velocities into velocities in the rectangular system of Galactic coordinates $(x,y,z)$  where the Sun is at the origin.  

Radial velocities (RV) are available in the literature for several of our analysed sources, which were obtained as part of the INfrared Spectra of Young Nebulous Clusters (IN-SYNC) ancillary program of the Apache Point Observatory Galactic Evolution Experiment (APOGEE) and published by \citet[][for IC~348]{Cottaar2014} and \citet[][for NGC~1333]{Foster2015}. \cite{Kounkel2018} recently reported on a new reduction of the APOGEE data taken in Orion, IC~348, NGC~1333 and other regions. We use here the data products from this recent reduction, since it implements an improved analysis of data variability.  These RVs were measured at multiple epochs with typical baselines of a few months. 
%\cite{Foster2015} give one single velocity and velocity uncertainty for each star, which are equal to the weighted mean and weighted mean  error, respectively, over all observed epochs. \cite{Cottaar2014}, on the other hand, give velocities taken at the individual  epochs.  
Thus, we compute for each star the average of all available radial velocities, after removing epochs where the signal-to-noise ratio of the associated spectrum is less than 20 and the best-fit effective temperature is less than 2400~K. As noted by \cite{Cottaar2015}, such epochs do not provide useful RVs and should be discarded in our analysis. For this analysis, we selected stars with rotational velocities in the range 5--150~km~s$^{-1}$ \citep{Cottaar2015}, RV uncertainties smaller than 2~km~s$^{-1}$ and excluded stars with very large proper motions, i.e.\ $>50$~km~s$^{-1}$.

Furthermore, we removed stars with variable RVs, since epoch-to-epoch variations would be induced by binaries with short periods whose orbital motions would introduce a velocity offset.  To look for strong radial velocity variability,  
%of sources in IC~348, 
we adopted the same procedure followed by \cite{Foster2015} in their own analysis of RVs. We computed the probability that the radial velocity is consistent with being constant, as estimated from the p-value that the $\chi^2=\sum\left(\rm{RV}_i-\mu\right)^2 / \sigma^2_i$ is larger than expected from chance, where $\rm{RV}_i$ is the radial velocity with uncertainty $\sigma_i$ in epoch $i$ and $\mu$ is the weighted mean over all epochs. Following \cite{Foster2015}, all sources with p-values smaller than $10^{-4}$ were excluded.  
%For the analysis of NGC~1333 stars, we adopt the flag indicating whether or not the source shows significant RV variability, which was provided along with the averaged data  (Column 15 of Table 3 in \citealt{Foster2015}).  
The number of sources used to investigate the kinematics of the clouds, after removing  the RV-variable sources is 133 in IC~348 and 31 in NGC~1333. Their radial velocities are given in Table \ref{tab:vel}. %We must note with mean values of km~s$^{-1}$

The  velocities  $(U,V,W)$  of each star relative to the $(x,y,z)$ reference system are listed in Table \ref{tab:vel}. These were transformed to $(u,v,w)$ LSR velocities  by subtracting the peculiar motion of the Sun, for which we use the values of the Solar motion  obtained by \cite{Schonrich_2010}: $U_0 = 11.1\pm0.7$~km~s$^{-1}$, $V_0 = 12.2\pm0.47$~km~s$^{-1}$ and $W_0 = 7.25\pm0.37$~km~s$^{-1}$. 

\begin{figure*}[!ht]
\begin{center}
\includegraphics[width=0.7\textwidth,angle=0]{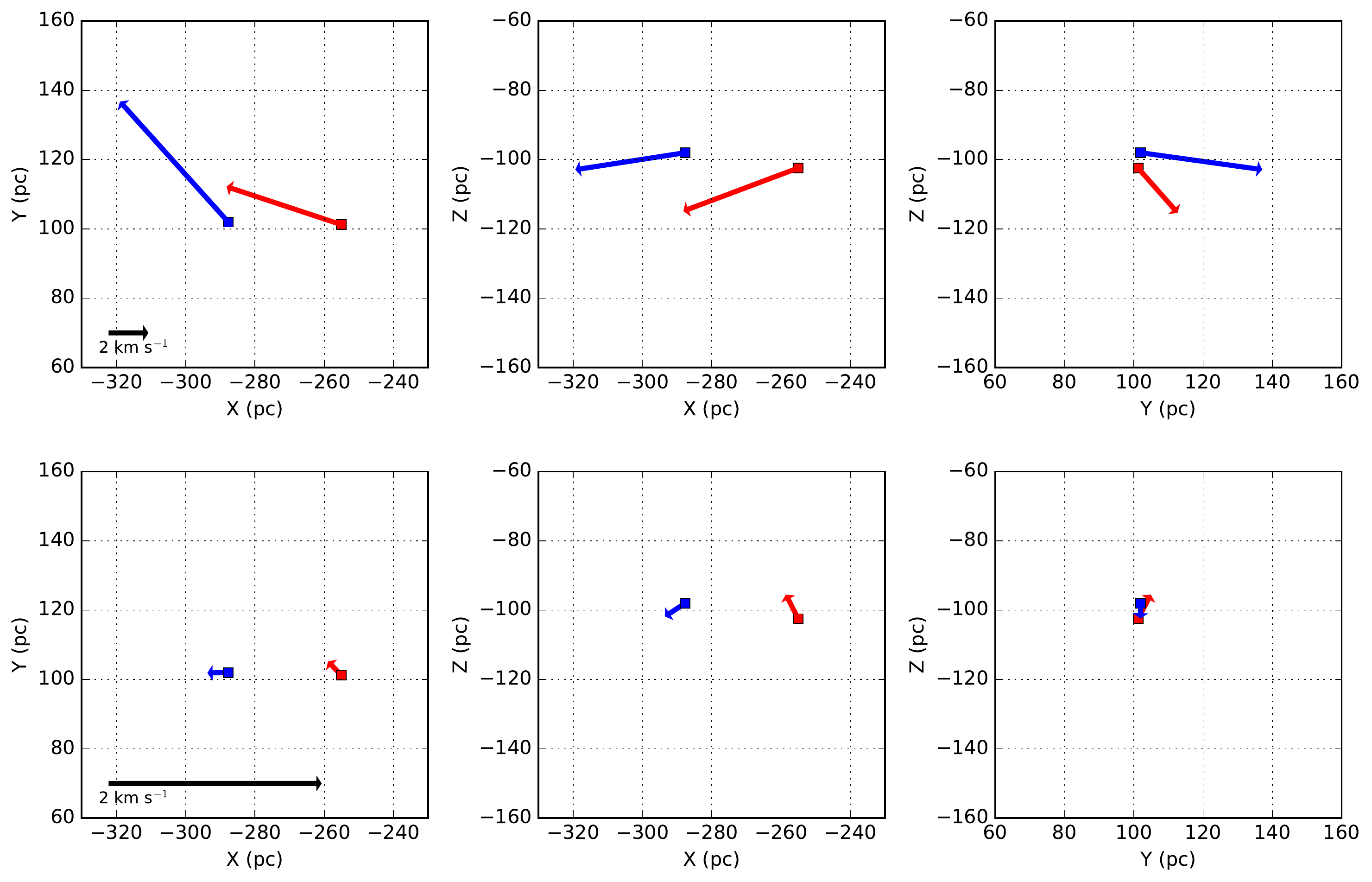}  
\end{center}
 \caption{Top: Mean of LSR velocities of stars in IC~348 (blue) and NGC~1333 (red) expressed in the rectangular system of Galactic coordinates. The origin of the arrows coincides with the mean of $(X,Y,Z)$ positions of the stars in each cluster. Bottom: Mean of the cross products $\mathbf{\hat{r}_\ast} \times \mathbf{\delta v_\ast}$ of stars in IC~348 (blue) and NGC~1333 (red). These vectors have been augmented by $6\times$ for better visualization. }
\label{fig:perseus-3d}
\end{figure*}

In the top panel of Figure \ref{fig:perseus-3d}, we show the projections of the mean LSR velocities $(\bar{u},\bar{v},\bar{w})$ as the blue and red arrows for IC~348 and NGC~1333, respectively.  We found $(\overline{U},\overline{V},\overline{W})_{\rm{IC~348}}=(-17.2\pm1.6, -6.2\pm1.1, -8.2\pm1.2)$ km~s$^{-1}$, $(\bar{u},\bar{v},\bar{w})_{\rm{IC~348}}=(-6.1\pm1.6, 6.8\pm1.1, -0.9\pm1.2)$ km~s$^{-1}$, $(\overline{U},\overline{V},\overline{W})_{\rm{NGC~1333}}=(-17.5\pm1.0, -10.9\pm1.4, -9.6\pm1.0)$ km~s$^{-1}$,  and $(\bar{u},\bar{v},\bar{w})_{\rm{NGC~1333}}=(-6.4\pm1.0, 2.1\pm1.4, -2.4\pm1.0)$ km~s$^{-1}$, where the quoted errors correspond to the standard deviation.  For the calculation of these spatial velocity components, we have  used the average distances derived in Section \ref{sec:structure}, because, as we have already pointed out, the individual parallaxes uncertainties are large (i.e.\ comparable to the parallax dispersion) that would broaden the velocity dispersion. The resulting 3D velocity dispersion is $\sigma = \sqrt{ \sigma_u^2 + \sigma_v^2 + \sigma_w^2 } = 2$ km~s$^{-1}$  for both clusters. 

There is a significant difference between the velocity vectors $(\overline{U},\overline{W},\overline{W})$ measured here and those measured for the Perseus OB2 association, which overlaps the Perseus Molecular cloud in the sky. On the basis of  Hipparcos proper motions, \cite{Belikov2002} found $(U,W,W)=(-12.7\pm1.6, -3.0\pm0.6, -0.9\pm0.8)$ km~s$^{-1}$ and a distance of $\sim300$~pc for the associattion. This is not surprising given that the Perseus OB2 association, with an age of $6$~Myr \citep{deZeeuw1999}, is older than both IC~348 and NGC~1333, and its dynamics has thus been affected by its interaction with the interstellar medium. 
%Clearly the clusters are moving in similar directions, which is consistent with the picture of both located at similar distances.
% in which IC 348 is embedded,

To calculate the expansion (or contraction) and rotation  velocities within each cluster, we use the same methodology as was used for the Taurus complex in \cite{Rivera_2015}. The expansion (or contraction) and rotation velocities are approximately given by the dot and cross products according to,

\begin{displaymath}
v_{\rm exp}=  \mathbf{\hat{r}_\ast} \cdot \mathbf{\delta v_\ast}, 
\end{displaymath}
\begin{displaymath}
v_{\rm rot}=  \mathbf{\hat{r}_\ast} \times \mathbf{\delta v_\ast},
\end{displaymath}

\noindent where $\mathbf{\hat{r}_\ast} = r_\ast/ |r_\ast|$ is the unit vector of the position of the star relative to the cluster center, and $\mathbf{\delta v_\ast}$ is the velocity of the star with respect to the cluster itself. 

These expansion and rotation velocities were computed for each star in our analyzed sample, and then we take the mean of each cluster to arrive at $v_{\rm exp,~IC~348}=-0.06$~km~s$^{-1}$ and $v_{\rm exp,~NGC~1333}=0.19$~km~s$^{-1}$. The resulting expansion velocities are very small compared with the velocity dispersion of 2~km~s$^{-1}$. This means that the stellar motions  in the radial direction do not seem to follow an expansion or contraction pattern.
% which is clearly visible in Figures \ref{} and \ref{}, where the individual motions do not show any organized patten   

The bottom panel of Figure \ref{fig:perseus-3d} shows the projection of the mean rotation velocities. $v_{\rm rot,~IC~348}=(-0.16, 0.0, -0.10)$ km~s$^{-1}$ and $v_{\rm rot,~NGC~1333}=(-0.10, 0.10, 0.19)$ km~s$^{-1}$. 
%The magnitudes of the rotation velocity in both clusters are also small in comparison with the velocity dispersions, which. 
These measurements suggest that the rotation velocity of both clusters is too small, if present at all. 

In IC~348, \cite{Cottaar2015} found a velocity gradient of  $0.024\pm0.013$~km~s$^{-1}$~arcmin$^{-1}$ due to a possible solid-body rotation of the cluster. Since the region under consideration has a size $\sim36$~arcmin (Figure \ref{fig:ic348-vel}), 
%Since the stars considered here are distributed over a region ,  one would expect to measure a 
this velocity gradient would imply a rotation velocity of $\sim0.9\pm0.5$~km~s$^{-1}$. Thus, the analysis presented here does not support the findings of \cite{Cottaar2015}.  It should be noted, moreover, that the statistical significance of that measurement is at the $1.8\sigma$ level.

%$(U,V,W)_{\rm{IC~348}}=(-16.3\pm1.3, -6.1\pm1.8, -8.9\pm1.6)$ km~s$^{-1}$, $(\bar{u},\bar{v},\bar{w})_{\rm{IC~348}}=(-5.2\pm1.3, 6.9\pm1.8, -1.6\pm1.6)$ km~s$^{-1}$, $(U,V,W)_{\rm{NGC~1333}}=(-16.2, -11.2, -10.6)$ km~s$^{-1}$,  and $(\bar{u},\bar{v},\bar{w})_{\rm{NGC~1333}}=(-5.1, 1.8, -3.4)$ km~s$^{-1}$

\clearpage

\section{Conclusions}\label{sec:conclusions}

We have performed multi-epoch VLBA observations of three objects embedded in IC~348 and one object in NGC~1333, which are located near opposite ends within the Perseus molecular cloud. From the astrometric fits of this sample we derived a mean distance of $321\pm10$~pc to IC~348, representing the most reliable distance determination to the eastern of Perseus. This distance is consistent with the mean of  Gaia DR2  parallaxes to a selected sample of known confirmed members of the cluster. The uncertainty on the mean of  Gaia  parallaxes is, however, 2.6 times larger than the VLBA uncertainty. The source detected with the VLBA in NGC~1333 is a close binary system, for which we derive preliminary orbital parameters.  Unfortunately, the VLBA data is not enough to provide a reliable distance for this specific source, and consequently, for the NGC~1333 cluster. Gaia parallaxes, on the other hand, yield a mean distance of $293\pm22$~pc. From these measurements, we conclude that the distance between the western and eastern edges of the clouds is about 30~pc in the direction of the line of sight. 

We use Gaia proper motions and radial velocities from the literature to derive the spatial velocities for a subset of cluster members. We derive the average spatial velocity vectors of IC~348 and NGC~1333, which are similar in magnitude and direction between them, but significantly different to the mean spatial motion of the Perseus OB2 association. We have estimated the expansion (or contraction) and rotation velocities of each cluster and found no clear evidence of such organized motions.  

%The main results of this paper are summarized in Table \ref{tab:properMotions}. 

\acknowledgements{G.N.O.-L.\ acknowledges support from the  von Humboldt Stiftung. P.A.B.G.\ acknowledges financial support from the S\~ao Paulo Research Foundation (FAPESP)  through grants 2013/04934-8 and 2015/14696-2. L.L. acknowledges the financial support of DGAPA, UNAM (project IN112417), and CONACyT, M\'exico. 

The Long Baseline Observatory is a facility of the National Science Foundation operated under cooperative agreement by Associated Universities, Inc. The National Radio Astronomy Observatory is a facility of the National Science Foundation operated under cooperative agreement by Associated Universities, Inc.  

This work has made use of data from the European Space Agency (ESA) mission {\it Gaia} (\url{https://www.cosmos.esa.int/gaia}), processed by the {\it Gaia} Data Processing and Analysis Consortium (DPAC, \url{https://www.cosmos.esa.int/web/gaia/dpac/consortium}). Funding for the DPAC has been provided by national institutions, in particular the institutions participating in the {\it Gaia} Multilateral Agreement.

Funding for SDSS-III has been provided by the Alfred P. Sloan Foundation, the Participating Institutions, the National Science Foundation, and the U.S. Department of Energy Office of Science. The SDSS-III web site is \url{http://www.sdss3.org/}.
SDSS-III is managed by the Astrophysical Research Consortium for the Participating Institutions of the SDSS-III Collaboration including the University of Arizona, the Brazilian Participation Group, Brookhaven National Laboratory, Carnegie Mellon University, University of Florida, the French Participation Group, the German Participation Group, Harvard University, the Instituto de Astrofisica de Canarias, the Michigan State/Notre Dame/JINA Participation Group, Johns Hopkins University, Lawrence Berkeley National Laboratory, Max Planck Institute for Astrophysics, Max Planck Institute for Extraterrestrial Physics, New Mexico State University, New York University, Ohio State University, Pennsylvania State University, University of Portsmouth, Princeton University, the Spanish Participation Group, University of Tokyo, University of Utah, Vanderbilt University, University of Virginia, University of Washington, and Yale University.
}

\clearpage
%\floattable
\begin{deluxetable}{llllc}
\tabletypesize{\scriptsize}
\tablecaption{VLBA observed epochs \label{tab:obs}}
\tablewidth{0pt}
%\tablecolumns{5}
\tablehead{ Project & Observation & \multicolumn{2}{c}{VLBA pointing positions} & Observed \\
code &  Date  & \colhead{R.A. ($\alpha_{2000}$)} &  \colhead{Decl. ($\delta_{2000}$)} & band \\
}\startdata
%%%%%
% BL175E0 &01 sep 2013 & 18:29:10.178 & +01:25:59.56 & C\\
BL175CD & 2012 sep 04 &  03:44:34.77 & 32:07:43.99 &  X \\
BL175CF & 2012 sep 07 & 03:45:00.92  & 32:04:19.03 & X \\
BL175AS & 2013 mar 22  & 03:44:34.77  & 32:07:43.99 & X \\
BL175AU & 2013 apr 18 & 03:45:00.92 & 32:04:19.03 & X \\
BL175H8 & 2014 apr 13 & 03:45:00.92  & 32:04:19.03 & X \\
BL175EF & 2014 sep 06 & 03:43:58.63  & 32:01:45.64 &  X \\
  & &03:44:21.89  &32:09:48.04  & \\
  & & 03:44:34.77 & 32:07:43.99 & \\ 
BL175CQ & 2014 sep 13 & 03:45:00.92  & 32:04:19.03 &  X \\
BL175EW & 2015 apr 26 & 03:45:00.92  & 32:04:19.03 & X \\
BL175HS & 2015 oct 20 & 03:45:00.92 & 32:04:19.03  & X \\
BL175HU & 2015 oct 24 & 03:43:58.63  & 32:01:45.64 & X \\
  & &03:44:21.89  &32:09:48.04  & \\
  & & 03:44:34.77 & 32:07:43.99 & \\ 
BL175I9 & 2016 apr 07 & 03:45:07.97 & 32:04:01.81 & C \\
BL175IB & 2016 apr 11 & 03:28:50.00 &31:30:00.00 &  C \\
  & & 03:29:03.00 & 31:22:00.00 & \\
  & & 03:29:20.00 & 31:14:00.00 & \\
BL175ID & 2016 apr 30 & 03:44:25.00 & 32:08:30.00  & C \\
 & & 03:44:45.00 & 32:17:00.00  & \\ 
BL175IP & 2016 aug 26 & 03:28:50.00 &31:30:00.00  & C \\
  & & 03:29:03.00 & 31:22:00.00 & \\
  & & 03:29:20.00 & 31:14:00.00 & \\
BL175J6 & 2016 aug 27 & 03:44:25.00 & 32:08:30.00 & C \\
 & & 03:44:45.00 & 32:17:00.00  & \\ 
BL175J1 & 2016 oct 11 & 03:45:07.97  & 32:04:01.81 & C \\
BL175JX & 2017 apr 24 & 03:45:07.97 & 32:04:01.81 & C \\
BL175KN & 2017 oct 06 &  03:45:07.97 & 32:04:01.81 & C \\
BL175KH & 2017 oct 07 & 03:28:46.49  & 31:29:43.50 & C \\
 & & 03:29:03.00   & 31:22:00.00 \\
 & & 03:29:20.00   & 31:14:00.00 \\ 
BL175KI & 2017 oct 13 & 03:44:25.00  & 32:08:30.00  & C \\
 & & 03:44:45.00   & 32:17:00.00  \\
BL175KY & 2018 mar 15 & 03:29:10.39   & 31:21:59.00  & C \\
 & & 03:44:32.59   & 32:08:42.35 \\ 
 & & 03:45:07.96   & 32:04:01.75  \\ 
%%%%%
\enddata
%\tablenotetext{a}{}
\end{deluxetable}
\clearpage

\clearpage
%\floattable
%\LongTables
\begin{deluxetable}{lcccc}
%\rotate
\tabletypesize{\scriptsize}
\tablewidth{0pt}
\tablecolumns{5}
\tablecaption{VLBA measured source positions \label{tab:positions}}
\tablehead{ Julian Day & $\alpha$ (J2000.0) & $\sigma_\alpha$ & $\delta$ (J2000.0) & $\sigma_\delta$    \\
%                                     & h:m:s &  seconds            &   $^\circ$:$'$:$''$ & arc-seconds          \\
}
\startdata
%%%%%   
\cutinhead{V913PER} % 
2456374.43416 & 3  44  32.58795488 & 0.00001138 & 32  8  42.372174 & 0.000296 \\ 
2457319.91111 & 3  44  32.58879683 & 0.00000899 & 32  8  42.354985 & 0.000213 \\ 
2457509.32628 & 3  44  32.58867044 & 0.00001174 & 32  8  42.348552 & 0.000277 \\ 
2457628.01584 & 3  44  32.58905351 & 0.00000866 & 32  8  42.348630 & 0.000213 \\ 
2458039.88818 & 3  44  32.58918298 & 0.00000718 & 32  8  42.340549 & 0.000184 \\ 
2458193.47921 & 3  44  32.58891352 & 0.00000484 & 32  8  42.335871 & 0.000118 \\ 
2458209.43523 & 3  44  32.58895319 & 0.00000232 & 32  8  42.335604 & 0.000064 \\ 
\cutinhead{V918PER} % 
first source: \\
2457319.91152 & 3  44  36.94228029 & 0.00000324 & 32  6  45.414751 & 0.000058 \\ 
2457509.32628 & 3  44  36.94223035 & 0.00001212 & 32  6  45.410445 & 0.000356 \\ 
2457628.01584 & 3  44  36.94268097 & 0.00001542 & 32  6  45.409728 & 0.000315 \\ 
2458039.88818 & 3  44  36.94304360 & 0.00000313 & 32  6  45.401953 & 0.000083 \\ 
2458209.43523 & 3  44  36.94289598 & 0.00000619 & 32  6  45.396791 & 0.000161 \\  
second source: \\
2456174.97590 & 3  44  36.96218751 & 0.00000269 & 32  6  44.952195 & 0.000068 \\ 
%2456374.43561 & 3  44  36.93868377 & 0.00001538 & 32  6  45.375757 & 0.000445 \\ 
\cutinhead{LRL11} % 
2456177.96799 & 3  45  07.96419667 & 0.00000301 & 32  4  01.790487 & 0.000101 \\ 
2456761.37584 & 3  45  07.96445493 & 0.00000820 & 32  4  01.776312 & 0.000194 \\ 
2456913.95729 & 3  45  07.96488174 & 0.00000705 & 32  4  01.774299 & 0.000246 \\ 
2457139.34132 & 3  45  07.96475367 & 0.00000248 & 32  4  01.766255 & 0.000075 \\ 
2457315.85692 & 3  45  07.96497603 & 0.00000207 & 32  4  01.763100 & 0.000052 \\ 
2457486.39137 & 3  45  07.96476855 & 0.00000214 & 32  4  01.756829 & 0.000062 \\ 
2457672.88009 & 3  45  07.96503320 & 0.00000930 & 32  4  01.754586 & 0.000217 \\ 
2457868.34505 & 3  45  07.96484623 & 0.00000337 & 32  4  01.748432 & 0.000126 \\ 
2458032.89439 & 3  45  07.96512694 & 0.00000938 & 32  4  01.746482 & 0.000227 \\ 
2458193.47921 & 3  45  07.96483889 & 0.00000819 & 32  4  01.742014 & 0.000237 \\ 
2458209.43523 & 3  45  07.96484894 & 0.00001005 & 32  4  01.741787 & 0.000255 \\ 
\cutinhead{2MASS J03291037+3121591} % 
first source: \\
2457490.44419 & 3  29  10.36879126 & 0.00001335 & 31  21  58.937104 & 0.000281 \\ 
2457627.07078 & 3  29  10.36934100 & 0.00001221 & 31  21  58.932573 & 0.000302 \\ 
2458033.95677 & 3  29  10.36911952 & 0.00000841 & 31  21  58.925023 & 0.000144 \\ 
second source: \\
2457490.44419 & 3  29  10.42062181 & 0.00000777 & 31  21  59.032952 & 0.000176 \\ 
2458033.95677 & 3  29  10.42184192 & 0.00001568 & 31  21  59.017736 & 0.000187 \\ 
2458193.47921 & 3  29  10.42173882 & 0.00000477 & 31  21  59.011139 & 0.000096 \\ 
2458209.43523 & 3  29  10.42183461 & 0.00002250 & 31  21  59.011471 & 0.000280 \\
%%%%%
\enddata
%\tablecomments{\dag multiple system}
%\tablenotetext{1}{GBS-VLA stands for Gould's Belt Very Large Array Distance Survey (Dzib et al.~ 2013).}
\end{deluxetable}
\clearpage

\clearpage
%\floattable
\begin{deluxetable}{lcccccccc}
%\rotate
\tabletypesize{\scriptsize}
\tablewidth{0pt}
%\tablecolumns{9}
\tablecaption{Astrometric solutions of the VLBA-detected sources and their counterparts in the Gaia DR2 catalog. \label{tab:gaia-prlx}}
\tablehead{   & \multicolumn{4}{c}{VLBA}  &  \multicolumn{4}{c}{Gaia}  \\ 
                    Name & parallax & distance & $\mu_\alpha\cos\delta$ & $\mu_\delta$     & parallax  & distance$^{a}$ & $\mu_\alpha\cos\delta$ & $\mu_\delta$    \\
                               & (mas)    &  (pc)       &(mas yr$^{-1}$)              & (mas yr$^{-1}$)  & (mas)       &  (pc)       & (mas yr$^{-1}$)             & (mas yr$^{-1}$) \\
                   (1)       &   (2)        &    (3)      &   (4)                                &  (5)                    & (6)            &  (7)         & (8)                                 & (9) \\
}
\startdata
%%%%%
 IRAS 03260+3111 & 3.136 $\pm$ 0.152& 319 $^{+16}_{-15}$ & 7.973 $\pm$ 0.083& -11.257 $\pm$ 0.121& 
                                                   --     & --  & -- & --  \\                                
 V913 Per & 3.119  $\pm$ 0.104 & 321 $^{+11}_{-10}$ & 2.458 $\pm$ 0.047 & -7.272 $\pm$ 0.133  & 
                     3.708 $\pm$ 0.262 & 270 $^{+21}_{-18}$ & 5.039$\pm$ 0.482 & -7.111 $\pm$ 0.281 \\
 V918 Per & 3.129 $\pm$ 0.512  & 320  $^{+63}_{-45}$ & 4.857 $\pm$ 0.335 & -6.750 $\pm$ 0.488  & 
                    1.852 $\pm$ 0.333  & 549 $^{+140}_{-94}$ &-3.321 $\pm$ 0.602 & -9.831 $\pm$ 0.439 \\ 
 LRL 11 &  2.680  $\pm$ 0.076 &  373 $^{+11}_{-10}$ & 2.37 $\pm$ 0.08  &  -8.271 $\pm$ 0.160 & 
                 2.665 $\pm$ 0.117 &  372 $^{+17}_{-16}$ &1.814 $\pm$ 0.214 & -9.807 $\pm$ 0.123 \\
%
%%%%%
\enddata
%\tablecomments{\dag multiple system}
\tablenotetext{a}{These values were taken from the distance catalog available from the Gaia TAP service of the Astronomisches Rechen Institut \citep[ARI;][]{Bailer-Jones2018}.  }
%\tablenotetext{2}{ }
\end{deluxetable}
\clearpage
\clearpage
%\floattable
\begin{deluxetable}{lcccccc}
\tabletypesize{\scriptsize}
\tablecaption{Orbital solutions for LRL~11. \label{tab:lrl11}}
\tablewidth{0pt}
%\tablecolumns{5}
\tablehead{  Parameter & Best-fit solution
}\startdata
$P$ (year)          & $6.3 \pm 0.4$          \\
$a$ (mas)          & $2.73 \pm 0.16$       \\
$T_P$ (JD)        & $2458942 \pm 208$ \\
$e$                    & $0.147 \pm 0.078$   \\
$\omega$ (deg) & $291.1 \pm 19.8$     \\
$i$ (deg)            & $49.1 \pm 6.8$         \\
$\Omega$ (deg)& $84.4 \pm 8.5$         
\enddata
%\tablenotetext{a}{Corrected for the parallax zero-point shift of $-30~\mu$as. }
\end{deluxetable}
\clearpage
\clearpage
%\floattable
\begin{deluxetable}{cccccccc}
%\rotate
\tabletypesize{\scriptsize}
\tablewidth{0pt}
\tablecolumns{5}
\tablecaption{Astrometric parameters and radial velocities of individual sources in IC~348 and NGC~1333. \label{tab:astro}}
\tablehead{ Star    & Parallax   & $\mu_\alpha\cos\delta$ & $\mu_\delta$      &   $V_r$                 \\
                              & (mas)       & (mas yr$^{-1}$)             & (mas yr$^{-1}$)  &    (km s$^{-1}$)     \\
                  (1)        &  (2)          &    (3)                              &   (4)                     &   (5)                      \\ }
\startdata
%%%%%
J03283651+3119289 & 3.17 $\pm$ 0.18 & 7.04 $\pm$ 0.22 & -10.14 $\pm$ 0.2 & 15.84 $\pm$ 0.3 \\
J03284407+3120528 & 3.4 $\pm$ 0.63 & 6.92 $\pm$ 0.98 & -9.91 $\pm$ 0.72 & 17.63 $\pm$ 0.38 \\
J03284764+3124061 & 2.79 $\pm$ 0.76 & 10.2 $\pm$ 1.24 & -10.87 $\pm$ 0.82 & 12.38 $\pm$ 0.37 \\
J03285119+3119548 & 3.18 $\pm$ 0.12 & 7.24 $\pm$ 0.15 & -9.55 $\pm$ 0.13 & 14.54 $\pm$ 0.11 \\
J03285216+3122453 & 3.31 $\pm$ 0.12 & 5.99 $\pm$ 0.14 & -10.01 $\pm$ 0.12 & 13.95 $\pm$ 0.14 \\
%
%%%%% 
\enddata
%\tablecomments{\dag multiple system}
\tablenotetext{1}{Only a portion of the table is shown here. The full table is available online.}
%\tablenotetext{2}{ }
\end{deluxetable}
\clearpage
\clearpage
%\floattable
\begin{deluxetable}{ccccccc}
\tabletypesize{\scriptsize}
\tablecaption{Derived properties for IC~348 and NGC~1333. \label{tab:properMotions}}
\tablewidth{0pt}
%\tablecolumns{5}
\tablehead{  & IC 348  & NGC1333 
}\startdata
$\varpi_{\rm VLBA}$~(mas) & $3.12\pm0.10$ & --  \\
$\varpi_{\rm Gaia}$~(mas) & $3.09\pm0.25$ & $3.38\pm0.26$ \\
$d_{\rm VLBA}$~(pc) & $321\pm10$ & -- \\ 
$d_{\rm Gaia}^a$~(pc) & $320^{+28}_{-24}$ & $293^{+24}_{-21}$ \\ 
$\overline{\mu}_\alpha\cos\delta$ (mas yr$^{-1}$) & 4.35 $\pm$ 0.03 & 7.34 $\pm$ 0.05 \\
$\overline{\mu}_\delta$ (mas yr$^{-1}$)                 & -6.76 $\pm$ 0.01 & -9.90 $\pm$ 0.03 \\
$\sigma_{\alpha}$  (mas yr$^{-1}$)                & 0.24 $\pm$ 0.03 & 0.92  $\pm$ 0.05 \\
$\sigma_{\delta}$   (mas yr$^{-1}$)                & 0.52 $\pm$ 0.01 & 0.60 $\pm$ 0.03 \\
$\sigma_{v_\alpha}$    (km s$^{-1}$)             & 0.36 $\pm$ 0.05 & 1.27 $\pm$ 0.07 \\
$\sigma_{v_\delta}$    (km s$^{-1}$)              & 0.80 $\pm$ 0.01 & 0.83 $\pm$ 0.04 \\
$(\overline{X},\overline{Y},\overline{Z})$~pc &  (-288,  102,  -98)  & (-255,   101, -102)\\
$(\overline{U},\overline{V},\overline{W})$ km~s$^{-1}$ & (-17.2,   -6.2,  -8.2) & (-17.5, -10.9, -9.6) \\ 
$(\overline{u},\overline{v},\overline{w})$ km~s$^{-1}$ & (-6.1,   6.8, -0.9) & (-6.4,  2.1, -2.4) \\
$(\sigma_{u},\sigma_{v},\sigma_{w})$ km~s$^{-1}$ & (1.6,  1.1,  1.2) & (1.0,  1.4,  1.0)   \\ 
$v_{\rm exp}$ (km~s$^{-1}$) & -0.06 & 0.19 \\
$\vec{v}_{\rm rot}$ (km~s$^{-1}$) & (-0.16,  0.0 , -0.10)  & (-0.10,  0.10, 0.19) \\
%
%%%%%
\enddata
\tablenotetext{a}{Corrected for the parallax zero-point shift of $-30~\mu$as. }
\end{deluxetable}
\clearpage
%
%
%\tablehead{ 
%\colhead{Cluster} & 
%\colhead{$\bar{\mu}_\alpha\cos\delta$} & 
%\colhead{$\bar{\mu}_\delta$}   & 
%\colhead{$\sigma_{\alpha}$}    & 
%\colhead{$\sigma_{\delta}$}     & 
%\colhead{$\sigma_{v_\alpha}$} & 
%\colhead{$\sigma_{v_\delta}$}  \\
%      & 
%      (mas yr$^{-1}$)  & 
%      (mas yr$^{-1}$)  & 
%      (mas yr$^{-1}$)  & 
%      (mas yr$^{-1}$) & 
%      (km s$^{-1}$)    & 
%      (km s$^{-1}$)     \\
%     (1) & 
%     (2) &    
%     (3) &  
%     (4) &   
%     (5) &   
%     (6) &   
%     (7)                    \\ 
%}

%%%%%
%IC 348      & 3.65 $\pm$ 0.02 & -6.76 $\pm$ 0.01 &  0.54 $\pm$ 0.02 & 0.58 $\pm$ 0.01  & 0.83 $\pm$ 0.03 & 0.89 $\pm$ 0.02  \\
%NGC1333 & 6.15 $\pm$ 0.05 & -9.90 $\pm$ 0.03 & 0.54  $\pm$ 0.05 & 0.60 $\pm$ 0.03  & 0.76 $\pm$ 0.07 & 0.84 $\pm$ 0.04 \\ 
%% 
\clearpage
\begin{deluxetable}{ccccccccccc}
%\rotate
\tabletypesize{\scriptsize}
\tablewidth{0pt}
\tablecolumns{10}
\tablecaption{Spatial velocities and positions of individual sources in IC~348 and NGC~1333. \label{tab:vel} }
\tablehead{ 2MASS   &  $U$       &    $V$           & $W$        & $u$               & $v$                & $w$  &   X  & Y   & Z                   \\
                   identifier  &  \multicolumn{3}{c}{(km s$^{-1}$)}   &  \multicolumn{3}{c}{(km s$^{-1}$)}      &  \multicolumn{3}{c}{(pc)}   \\
                   (1)           &   (2)        &    (3)              &   (4)        &  (5)                & (6)                  &  (7)   & (8) &(9) & (10)                \\
}
\startdata
J03283651+3119289 & -18.06 & -10.32 & -10.59 & -6.96 & 2.68 & -3.34 & -254.69 & 101.72 & -102.91 \\
J03284407+3120528 & -19.56 & -9.37 & -11.05 & -8.46 & 3.63 & -3.8 & -254.77 & 101.71 & -102.75 \\
J03284764+3124061 & -17.2 & -15.09 & -7.63 & -6.1 & -2.09 & -0.38 & -254.81 & 101.85 & -102.5 \\
J03285119+3119548 & -17.11 & -10.43 & -9.34 & -6.01 & 2.57 & -2.09 & -254.83 & 101.57 & -102.73 \\
J03285216+3122453 & -15.72 & -9.9 & -10.58 & -4.62 & 3.1 & -3.33 & -254.84 & 101.72 & -102.54 \\
\enddata
%\tablecomments{\dag multiple system}
\tablenotetext{1}{Only a portion of the table is shown here. The full table is available online.}
%\tablenotetext{2}{ }
\end{deluxetable}
\clearpage
\clearpage
%\floattable
\begin{deluxetable}{cccccccc}
%\rotate
\tabletypesize{\scriptsize}
\tablewidth{0pt}
\tablecolumns{6}
\tablecaption{Gaia parallaxes and proper motions of YSO candidates outside NGC~1333 and IC~348 \label{tab:other-gaia}}
\tablehead{ Spitzer source   & Parallax & $\mu_\alpha\cos\delta$ & $\mu_\delta$          \\
                   name                 & (mas)     & (mas yr$^{-1}$)              & (mas yr$^{-1}$)       \\
                   (1)                     &   (2)        &    (3)                                &   (4)                         \\
}
\startdata
%%%%%
J032519.5+303424 & 0.68 $\pm$ 0.43 & 1.21 $\pm$ 0.72 & 0.08 $\pm$ 0.47 \\
J032835.0+302009 & 1.9 $\pm$ 1.98 & 5.7 $\pm$ 2.83 & -9.82 $\pm$ 1.94 \\
J032842.4+302953 & 3.62 $\pm$ 0.14 & 6.36 $\pm$ 0.26 & -9.75 $\pm$ 0.15 \\
J033052.5+305417 & 3.17 $\pm$ 0.3 & 7.09 $\pm$ 0.37 & -7.79 $\pm$ 0.31 \\
J033118.3+304939 & 3.32 $\pm$ 0.08 & 7.27 $\pm$ 0.12 & -7.8 $\pm$ 0.09 \\
J033120.1+304917 & 3.9 $\pm$ 0.24 & 7.86 $\pm$ 0.35 & -8.25 $\pm$ 0.24 \\
J033142.4+310624 & 3.67 $\pm$ 0.17 & 7.91 $\pm$ 0.25 & -6.56 $\pm$ 0.17 \\
J033241.6+311044 & 4.05 $\pm$ 0.84 & 7.58 $\pm$ 1.59 & -8.63 $\pm$ 0.84 \\
J033241.7+311046 & 2.78 $\pm$ 0.24 & 7.65 $\pm$ 0.49 & -7.65 $\pm$ 0.26 \\
J033312.8+312124 & 2.2 $\pm$ 0.57 & 6.12 $\pm$ 0.71 & -7.93 $\pm$ 0.6 \\
J033330.4+311050 & 2.62 $\pm$ 0.29 & -2.0 $\pm$ 0.52 & -3.29 $\pm$ 0.34 \\
J033346.9+305350 & 3.67 $\pm$ 0.2 & 10.92 $\pm$ 0.31 & -12.58 $\pm$ 0.2 \\
J033915.8+312430 & 2.37 $\pm$ 0.7 & 7.58 $\pm$ 1.35 & -6.49 $\pm$ 0.81 \\
%%%%%%
\enddata
%\tablecomments{\dag multiple system}
%\tablenotetext{1}{ }
%\tablenotetext{2}{ }
\end{deluxetable}
\clearpage

\bibliographystyle{aasjournal}
\bibliography{ms.bib}

\end{document}